\documentclass[12pt,eadjoint tfnpsf]{article}

\catcode`\@=11
\@addtoreset{equation}{section}

\global\arraycolsep=1pt

\usepackage{amsbsy,amssymb,latexsym,amsfonts, amsmath}
\usepackage{mathrsfs}
\usepackage{graphicx}

\newcommand{\be}{\begin{equation}}
\newcommand{\ee}{\end{equation}}
\newcommand{\bea}{\begin{eqnarray}}
\newcommand{\eea}{\end{eqnarray}}

\newcommand{\CB}{{\mathcal B}}
\newcommand{\CC}{{\mathcal C}}

\newcommand{\CN}{{\mathcal N}}
\newcommand{\CO}{{\mathcal O}}

\newcommand{\CW}{{\mathcal W}}

\newcommand{\1}{\mbox{1}\hspace{-0.25em}\mbox{l}}

\newcommand{\complex}{{\mathbb {C}}} 
\newcommand{\zet}{{\mathbb{ Z}}} 

\begin{document}
%
%
\begin{titlepage}

\begin{flushright}
\normalsize
~~~~
SISSA  23/2012/EP-FM\\
\end{flushright}

\vspace{70pt}

\begin{center}
{\huge ${\cal N}=2$ gauge theories on toric singularities,}\\
\vspace{12pt}
{\huge blow-up formulae and $W$-algebrae}\\
\end{center}

\vspace{25pt}

\begin{center}
{
Giulio Bonelli$^{\heartsuit\spadesuit}$, 
Kazunobu Maruyoshi$^\heartsuit$, 
Alessandro Tanzini$^\heartsuit$
 and 
Futoshi Yagi$^\heartsuit$
}\\
%
\vspace{13pt}
%
${}^\heartsuit${\it SISSA and INFN, Sezione di Trieste, via Bonomea 265, 34136 Trieste, Italy}\\
${}^\spadesuit$ {\it I.C.T.P. -- Strada Costiera 11, 34014 Trieste, Italy}\\
\end{center}
%
\vspace{17pt}
\begin{center}
Abstract\\
\end{center}
We compute the Nekrasov partition function of gauge theories on the (resolved) toric singularities 
$\mathbb{C}^2/\Gamma$ in terms of blow-up formulae. 
We discuss the expansion of the partition function in the $\epsilon_1,\epsilon_2\to 0$ limit
along with its modular properties and how to derive them from the M-theory perspective. 
On the two-dimensional conformal field theory side, our results can be interpreted in terms of representations of the direct sum of Heisenberg plus $W_N$-algebrae with suitable central charges, which can be computed from the fan of the resolved toric variety.
We provide a check of this correspondence by computing the central charge of the two-dimensional theory from 
the anomaly polynomial of M5-brane theory. Upon using the AGT correspondence our results provide a candidate for 
the conformal blocks and three-point functions of a class of the two-dimensional CFTs 
which includes parafermionic theories.  



\vfill

\vfill

\vspace{50pt}
\hrule height 0.05mm depth 0.1mm width 70mm
\vspace{10pt}
{\small e-mails: bonelli, maruyosh, tanzini, fyagi@sissa.it}

\setcounter{footnote}{0}
\renewcommand{\thefootnote}{\arabic{footnote}}

\end{titlepage}

\section{Introduction}
\label{sec:intro}
  ${\cal N}=2$ gauge theories in four-dimensions provide a seemingly inexhaustible source of results
  in theoretical physics and geometry. 
  The partition function of the gauge theory on the so-called Omega background \cite{Nekrasov:2002qd} 
  with parameters $\epsilon_{1,2}$ gives the exact answer not only to the prepotential of the theory, 
  summing over all the instanton contributions, in the leading behavior $\epsilon_{1,2} \rightarrow 0$, 
  but also to the gravitational corrections included in finite $\epsilon$ terms.
  The introduction of the Omega background also enables us to concatenate $\CN=2$ theories 
  with two-dimensional conformal field theories (CFT) \cite{Alday:2009aq} 
  and quantum integrable systems \cite{Nekrasov:2009rc}.
  
  In this paper we present a systematic study of these theories 
  on the minimal resolution $X_{p,q}$ of the toric singularities $\mathbb{C}^2/\Gamma_{p,q}$ 
  with $\Gamma_{p,q}\subset U(2)$ a finite group.
  We compute the full Nekrasov partition function by reducing the problem 
  to a diagrammatic algorithm related to the fan of the toric variety.
  We then analyze the geometry of the low-energy effective action and its modular properties, 
  by studying the $\epsilon_{1,2} \to 0$ limit and by elucidating their M-theory origin. 
  We find that there are two types of contributions, being related respectively 
  to the regular and irreducible representations of $\Gamma_{p,q}$. 
  The former is encoded in the Seiberg-Witten curve $\Sigma$ of the gauge theory on the flat space, 
  up to an overall volume factor. 
  The twisted sector is encoded in suitable modular functions whose lattice is determined 
  by the intersection matrix of the Hirzebruch-Jung resolution of the singularity.
  Indeed we show that the twisted sector contribution can be written in term of blow-up equations 
  which generalize the ones of \cite{FS,Moore:1997pc,Marino:1998bm} to the blow-up of {\it singular} points. 

  Let us observe that in general
  ${\cal N}=2$ theories can be formulated on any differentiable
  four-manifold by using a twisting procedure \cite{Witten:1988ze}. 
  A cautionary remark is in order here: when $\Gamma$ is a Kleinian subgroup of $SU(2)$
  the resolved manifold is an ALE space, which displays an hyper-kahler structure. 
  Then the original and the twisted theory share the same energy-momentum tensor and are physically equivalent. 
  For more general $\Gamma_{p,q}$ subgroups the resolved manifolds are only Kahler and preserve 
  half of the supersymmetric charges with respect to the ALE case. 
  The energy-momentum tensor is in this case substantially modified and only
  the twisted version of the theory preserves supersymmetry. 
  
  In section \ref{sec:gauge}, the full partition function will be given
  and expanded in $\epsilon_{1,2}$ to see the geometric properties.
  Remarkably, the partition function is a nested product of $\mathbb{C}^2$ partition functions 
  with arguments shifted according to the $X_{p,q}$ geometry.
  By studying the leading orders in the expansion we show that the low-energy effective action 
  is encoded in the Seiberg-Witten curve and the intersection
  matrix of the resolved toric variety. 
  We also discuss next to leading orders including gravitational corrections and the Nekrasov-Shatashvili 
  limit of the full partition function. 
  
  In section \ref{sec:instanton}, we consider the Nekrasov partition function on $X_{p,q}$ 
  at classical, one loop, and instanton level, separately and
  we relate them with the blowup formula at each level.
  
  In section \ref{sec:M}, we provide a description of the system in terms of M5-branes 
  on $X_{p,q}\times {\cal C}$ where ${\cal C}$ is a punctured Riemann surface. 
  We discuss how the modular properties of the low-energy effective action are captured 
  by the generalized elliptic index of the M5-brane system in the far infrared, 
  where the system reduces to a single M5-brane wrapping the Seiberg-Witten curve $\Sigma$. 
  
  The M5-brane picture can be also used to gain some insights on the corresponding two-dimensional CFT
  {\it \`a la} AGT \cite{Alday:2009aq}. 
  Indeed, our result for the full Nekrasov partition function indicates the emergence 
  of representations of the algebra ${\cal A}_N (X_{p,q})$ 
  given by direct sum of Heisenberg plus $W_N$ algebrae with suitable central charges 
  which can be computed from the fan of the toric variety, see \eqref{copies}. 
  On the other hand, the central charge of the candidate two-dimensional CFT can be computed from
  the anomaly polynomial of the $N$ M5-branes wrapping ${\cal C}$ via equivariant integration 
  over the four-dimensional space \cite{Bonelli:2009zp,Alday:2009qq,Nishioka:2011jk}. 
  In section \ref{sec:CFT}, we check that this reproduces the overall central charge of the algebra ${\cal A}_N(X_{p,q})$.
  
  We conclude in section \ref{sec:conclusion} with various discussions.
  In the Appendices, we review the properties of $X_{p,q}$ spaces,
  collect useful formulae which are needed in the computation and expansion of the partition function
  and
  we explicitly give the first terms in the expansion of the instanton sum for the $U(2)$ SYM theory on $A_2$ ALE.

\section{${\cal N}=2$ gauge theories on toric singularities}
\label{sec:gauge}

The partition function of ${\cal N}=2$ $U(N)$ gauge theories can be computed via equivariant localization
methods on a general toric manifold $X$. 
Indeed, one exploits the $(\complex^*)^{2+N}$ action on the moduli space of instantons on $X$, 
where $(\complex^*)^2$ is the lift of space-time automorphisms of $X$ to the instanton moduli space 
while $(\complex^*)^N$ is the complexification of the Cartan torus of the $U(N)$ gauge symmetry.  
For compact toric rational surfaces this was studied in \cite{Gasparim:2008ri}. 
 In this section we discuss open varieties focusing on the most general toric singularity
$\mathbb{C}^2/\Gamma_{p,q}$, with $\Gamma_{p,q}\subset U(2)$ a finite group acting on local coordinates as 
$z_1\to e^{2\pi i/p}z_1$ and $z_2\to e^{2\pi i q/p} z_2$, with $(p,q)$ being coprime
and $q<p$. More precisely we consider the minimal resolution of this singularity 
$X_{p,q}=\widetilde{\mathbb{C}^2/\Gamma_{p,q}}$, known as Hirzebruch-Jung
resolution -- see Appendix \ref{HJ} for details.
  The $\mathcal{N}=4$ partition function for these geometries was calculated 
  in \cite{Fucito:2006kn,Griguolo:2006kp,Bruzzo:2009uc,Cirafici:2009ga}. 

The general procedure to compute the ${\cal N}=2$ Nekrasov partition function 
is the following: any toric variety is described in terms
of a fan encoding its patching structure as a complex manifold.
In each patch the computation of the Nekrasov partition function reduces to the standard one in $\complex^2$
spanned by suitable variables which provide a basis of invariants of the orbifold action $\Gamma_{p,q}$.
One thus obtains a diagrammatic algorithm which computes the full partition function
from the weights of the $(\complex^*)^{N+2}$-torus action in each patch.
Indeed, as it has been suggested in \cite{Nekrasov:2003vi} and then shown 
in \cite{Nakajima:2003uh,Bonelli:2011jx,Bonelli:2011kv}
for the blown-up $\mathbb{P}^2$ and $\mathcal{O}_{\mathbb{P}^1}(-2)$ cases, 
this description is particularly simple when one considers the full ${\cal N}=2$
partition function including the classical and perturbative contributions. The full partition 
function on the resolved toric singularity is simply given by the intertwined product of the full partition functions in 
each patch. 
More precisely, we propose that the ${\cal N}=2$ full Nekrasov partition function on $X_{p,q}$
is given by the blowup formula
    \bea
    Z^{X_{p,q}}_{{\rm full}}(\vec{a},\epsilon_1,\epsilon_2)=\sum_{\{\vec k^{(\ell)}\}} \prod_{\ell=0}^{L-1} 
    Z^{\complex^2}_{{\rm full}}(\epsilon_1^{(\ell)},\epsilon_2^{(\ell)}, \vec{a}^{(\ell)} )
    \xi_{\ell}^{c_{1}^{(\ell)}},
    \label{full}
    \eea
  where $\vec a = \{ a_\alpha \}, \alpha=1,\ldots,N$ are the vev's of the scalar field 
  of the ${\cal N}=2$ vector multiplet,
 \bea
 a^{(\ell)}_{\alpha} 
     = a_{\alpha} 
         + k^{(\ell+1)}_{\alpha} \epsilon_1^{(\ell)} 
         + k^{(\ell)}_{\alpha} \epsilon_2^{(\ell)}
\label{shift} 
\eea  
and $k^{(0)}=k^{(L)}=0$.

The above formula (\ref{full}) can be obtained as follows. 
As reviewed in Appendix \ref{HJ}, the $X_{p,q}$ variety 
is described in terms of $L$ patches with local coordinates described in (\ref{t}). These local coordinates
transform under the $(\complex^*)^2$ torus action with weights $\left(\epsilon_1^{(\ell)},\epsilon_2^{(\ell)}\right)$
whose explicit expression is given in (\ref{w-bis}). 
The fixed point data on $X_{p,q}$ are described in terms of a collection of Young tableaux $\{\vec Y_\ell \}$,
and of rational numbers $\{\vec k^{(\ell)} \}$ $\ell=0, \ldots,L-1$ describing respectively the $(\complex^*)^{N+2}$-invariant 
point-like instantons in each patch and the magnetic fluxes of the gauge field on the blown-up spheres
which correspond to the first Chern class $c_1(E)$ of the gauge bundle $E$. 
  More explicitly, the homology decomposition of the $c_1$ reads
\be
c_1 = -\sum_{\alpha,\ell} k_{\alpha}^{(\ell)} \Delta_\ell,
\ee
where $\{\Delta_\ell\}$ is a basis of $H_2(X_{p,q})$. 
Since the $c_1$ has an integer decomposition in the dual cohomology basis
 \be
c_1 = \sum_{\alpha,\ell} u_{(\ell) \alpha} \omega_\ell  \ \ , \vec u_{(\ell)} \in\zet^{N(L-1)},
\ee 
where $\omega_\ell$ is a basis of $H^{2-}(X_{p,q})$ normalized by the condition $\int_{\Delta_\ell}\omega_n=\delta_n^\ell$, 
we get that $\vec k = C^{-1} \vec u$ 
where ${\cal I}_{p,q}\equiv -C$ is the intersection form of the resolved $X_{p,q}$ variety 
displayed in the Appendix \ref{HJ}, eq.(\ref{C}).
Therefore, the lattice summation in \eqref{full} is 
$\{\vec k^{(\ell)}\} \in (\1_{N} \otimes C^{-1}) \mathbb{Z}^{N(L-1)}$.
  
  Note also that we have multiplied $\xi_{\ell}$ factors 
  in order to keep track of the first Chern classes $c_1^{(\ell)}$ of the gauge bundle
    \bea
    c_1^{(\ell)} 
     =     \sum_{\alpha=1}^N u_{(\ell) \, \alpha}
     =     \sum_{\alpha=1}^N \sum_{m=1}^{L-1} C_{\ell m} k^{(m)}_{\alpha}.
    \eea 
  In other words, we are considering the expectation value 
  $Z_{\rm full}^{X_{p,q}} = \left< e^{\frac{1}{2\pi}
  \sum_\ell z_{\ell} \int Tr(F) \wedge \omega_{(\ell)}} \right>$
  with $\xi_{\ell} = e^{(C^{-1}z)^{(\ell)}}$, rather than the partition function.

The shift in the Cartan parameters (\ref{shift}) can be computed by the patch-to-patch 
relative shift of the $\left(\complex^*\right)^N$ weights which is induced by the non-trivial magnetic flux
of the gauge field on the blown-up spheres as explained in the following.
We denote the $\alpha$-th gauge field on the north patch as $(A_N)_{(\ell)\alpha}$ 
while that on the south patch as $(A_S)_{(\ell)\alpha}$ in the 
$\ell$-th blown up sphere ($\ell=1,\cdots,L-1$).
At the equator, they coincide up to the gauge transformation
\begin{align}
(A_N)_{(\ell)\alpha} 
 =   (A_S)_{(\ell)\alpha} + \partial_{\phi} \psi_{(\ell)\alpha}(\phi)
 \label{ANAS}
\end{align}
where $\phi$ is the coordinate along the equator.
When we go around the equator,
the phase is identified up to a multiple of $2\pi$:
\bea
\psi_{(\ell)\alpha}(\phi+2\pi) = \psi_{(\ell)\alpha}(\phi) - 2 \pi u_{(\ell)\alpha}
\label{ppu}
\eea
with $u_{(\ell) \alpha} \in \zet$ being the magnetic flux through the blown-up sphere $\Delta_{\ell}$
\bea 
\frac{1}{2\pi}\int_{\Delta_{\ell}} F_{(\ell)\alpha} = {u}_{(\ell) \alpha}.
\eea

  According to (\ref{ANAS}) and (\ref{ppu}) the non-trivial flux through the $\ell$-th blown-up sphere 
  modifies the relative weights of the gauge $U(1)^N$ action $a^{(\ell)}_{\alpha}$ and
  $a^{(\ell-1)}_{\alpha}$ by
  $a^{(\ell)}_{\alpha} = a^{(\ell-1)}_{\alpha} - u_{(\ell)\alpha} \epsilon_1^{(\ell)}$.
 In order to get \eqref{shift} we parametrize the weights of the Cartan action
 as seen from the first and last patches and impose consistency. 
 The difference between $a^{(0)}_{\alpha}$ and $a_{\alpha}$
 is proportional to $\epsilon^{(0)}_1 = p\epsilon_1$,
 \bea
 a^{(\ell)}_{\alpha} = a_{\alpha} + x \epsilon_1 - \sum_{m=1}^{\ell} u_{(m)\alpha} \epsilon_1^{(m)}.
 \eea
 Furthermore, the difference between $a^{(L-1)}_{\alpha}$ and $a_{\alpha}$
 is proportional to $\epsilon_2^{(L-1)} = p\epsilon_2$. Henceforth, we obtain the condition
 \bea
 a^{(L-1)}_{\alpha} + y \epsilon_2 = a_{\alpha}
 \,\, \Rightarrow \,\,
 x \epsilon_1 - \sum_{m=1}^{L-1} u_{(m)\alpha} \epsilon_1^{(m)} + y \epsilon_2 = 0,
 \eea
 which determine the coefficients $x$ and $y$ as 
 \bea
 x = - \sum_{\ell=1}^{L-1} (p q_{\ell} - q p_{\ell}) u_{(\ell)\alpha},
 \qquad
 y = - \sum_{\ell=1}^{L-1} p_{\ell} u_{(\ell)\alpha} .
 \label{xy}
 \eea
 By using this result one gets \eqref{shift} as explained in detail at the end of Appendix B.

Due to the asymmetric nature of the $\Gamma_{p,q}$ orbifold, the usual symmetry $\epsilon_1\leftrightarrow\epsilon_2$ appearing in the 
flat $\mathbb{C}^2$ case is now replaced by the invariance of the full partition function under the simultaneous
exchange $q\leftrightarrow p_{L-1}$. This in turn is the reversal of the continuous fraction $[e_1,\ldots,e_{L-1}]\leftrightarrow
[e_{L-1},\ldots,e_1]$ and pictorially corresponds to reverse the order of the chain of blown-up spheres.

\subsection{Blowup formulae and theta functions}
\label{subsec:blowup}
  In this subsection, we discuss the behavior of the Nekrasov partition function on $X_{p,q}$ 
  in the limit $\epsilon_{1,2} \to 0$.  
  As we will show, this enables us to uncover the modular properties 
  of the ${\cal N}=2$ partition function and to derive a generalization of the blow-up equations 
  for the Donaldson polynomials \cite{FS,Marino:1998bm} to the case of toric singularities.
  
  First of all, we expand the full partition function on $\mathbb{C}^2$ as
    \bea
    & & - \epsilon_1 \epsilon_2 \ln Z^{\mathbb{C}^2}_{{\rm full}} (\epsilon_1, \epsilon_2, \vec{a}, q)
     \equiv
          F_{\Omega}(\epsilon_{1}, \epsilon_{2}, \vec{a}, q)
          \nonumber \\
    & &   \equiv F_0(\vec{a}, q) + (\epsilon_1 + \epsilon_2) H(\vec{a},q) + \epsilon_1 \epsilon_2 F_1(\vec{a}, q)
        + (\epsilon_1 + \epsilon_2)^2 G(\vec{a}, q) + \mathcal{O} (\epsilon^3).
          \label{expansion}
    \eea
  Note that these include the classical and the perturbative part.
  The leading part $F_0(\vec{a}, q)$ is the prepotential 
  and related to the IR effective gauge coupling constant as 
    \bea
    \left( \tau^{\mathbb{C}^2}_{\rm eff} \right)^{\alpha\beta} 
     \equiv    \partial^{\alpha} \partial^{\beta} F_0(\vec{a}, q).
    \eea
  By substituting the expansion (\ref{expansion}) into each of the $Z^{\mathbb{C}^2}$ factors
  in the blowup formula (\ref{full}), we obtain
    \begin{align}
    & - \ln Z^{\mathbb{C}^2}_{{\rm full}}
    (\epsilon_1^{(\ell)},\epsilon_2^{(\ell)}, \vec{a} +\epsilon_1^{(\ell)}\vec{k}^{(\ell+1)} 
    +\epsilon_2^{(\ell)}\vec{k}^{(\ell)})
    \cr
     = &   \frac{1}{\epsilon_1^{(\ell)}\epsilon_2^{(\ell)}} F_{0} (\vec{a}, q) 
           + \frac{\epsilon_1^{(\ell)} + \epsilon_2^{(\ell)}}{\epsilon_1^{(\ell)}\epsilon_2^{(\ell)}} H(\vec{a},q) 
       + \frac{\epsilon_1^{(\ell)} k^{(\ell+1)}_{\alpha} + \epsilon_2^{(\ell)} k^{(\ell)}_{\alpha}}%
       {\epsilon_1^{(\ell)}\epsilon_2^{(\ell)}}
       \partial^{\alpha} F_0(\vec{a}, q)  
        \cr
       & + F_1(\vec{a},q) 
        + \frac{(\epsilon_1^{(\ell)} + \epsilon_2^{(\ell)})^2}{\epsilon_1^{(\ell)}\epsilon_2^{(\ell)}} G(\vec{a},q) 
         + \frac{(\epsilon_1^{(\ell)} + \epsilon_2^{(\ell)})
        (\epsilon_1^{(\ell)} k^{(\ell+1)}_{\alpha} + \epsilon_2^{(\ell)} k^{(\ell)}_{\alpha} )}%
        {\epsilon_1^{(\ell)} \epsilon_2^{(\ell)} } 
        \partial^{\alpha}H(\vec{a}, q) 
       \cr
      & + \left( \tau^{\mathbb{C}^2}_{\rm eff} \right)^{\alpha\beta} 
        \frac{(\epsilon_1^{(\ell)} k^{(\ell+1)}_{\alpha} + \epsilon_2^{(\ell)} k^{(\ell)}_{\alpha} )
        (\epsilon_1^{(\ell)} k^{(\ell+1)}_{\beta} + \epsilon_2^{(\ell)} k^{(\ell)}_{\beta} )}%
        {\epsilon_1^{(\ell)} \epsilon_2^{(\ell)} } 
+ \mathcal{O} (\epsilon^3).
    \label{process1}
    \end{align}
  Then, we need to sum over $\ell$. 
  By using the identities in Appendix \ref{sec:sum}
  and by comparing with original partition function on $\mathbb{C}^2$,
  we obtain the following result up to terms of order one in $\epsilon_1,\epsilon_2$ in the exponential
    \bea
    & &    Z_{\rm full}^{X_{p,q}}(\epsilon_1, \epsilon_2, \vec{a}, q) 
     \simeq     \left( Z_{\rm full}^{\mathbb{C}^2} (\epsilon_1, \epsilon_2, \vec{a}, q) \right)^{\frac{1}{p}}
           \exp \left[ \left( - L + \frac{1}{p} \right) F_1(\vec{a},q) + N_{G} G(\vec{a},q) \right]
           \label{pippo} \\
    & &    ~~~~~~~
           \times \sum_{\{ \vec{k}^{(\ell)} \}} \xi_{\ell}^{c_{1}^{(\ell)}} \exp \left[ 
           \frac{1}{2} \sum_{\ell,m=1}^{L-1} \sum_{\alpha, \beta =1}^{N} 
           k_{\alpha}^{(\ell)} C_{\ell m} 
           \left( \tau_{\rm eff}^{\mathbb{C}^2} \right) ^{\alpha\beta}
           k_{\beta}^{(m)} 
         + \sum_{\alpha=1}^N \sum_{\ell=1}^{L-1}
           (e_{\ell}-2) k_{\alpha}^{(\ell)} 
           \partial_{\alpha} H
           \right],
           \nonumber
    \eea
  where 
    \bea
    N_{G}
     =     \frac{p_{L-1}+q+2}{p} - 2L + \sum_{\ell=1}^{L-1} e_{\ell}.
           \label{NG}
    \eea
  $p_{L-1}$ and $e_{\ell}$ are defined in Appendix \ref{HJ}.
  
  The above result shows that the leading term in the $\epsilon_{1,2} \rightarrow 0$ limit 
  is simply the same as the gauge theory prepotential on $\mathbb{C}^2$ up to a factor $p$
    \bea
    Z_{\rm full}^{X_{p,q}}
     =     \exp \left( - \frac{F_0}{p \epsilon_1 \epsilon_2} + \ldots \right).
    \eea
  This can be interpreted as follows:
  the low energy effective theory has a sector which is described 
  by the very same Seiberg-Witten curve $\Sigma$ and differential $\lambda_{{\rm SW}}$ as the flat
  space. However, the volume factor is rescaled by the order of the quotient group. This sector
  corresponds to point-like instantons sitting in the regular representation of $\Gamma_{p,q}$.
  These probe the whole quotient space $X_{p,q}$ and their contribution is weighted by the
  equivariant volume. Besides this sector, there is also the one of instantons in the irreducible
  representation of the finite group. These are stuck at the invariant loci of the orbifold action,
  namely on the blown-up spheres, and as such their contribution is independent on $\epsilon_1,\epsilon_2$.
  This contribution is fully characterized by the intersection matrix of $X_{p,q}$ as displayed in the second
  line of (\ref{pippo}). The subleading terms in the first and second lines of \eqref{pippo} represent 
  the gravitational couplings.
  
  One can check that the same behavior holds in the NS limit $\epsilon_2\to 0$ with $\epsilon_1$ finite 
  \cite{Nekrasov:2009rc}.
  In particular, regular instantons contribute as follows
    \bea
    \CW^{X_{p,q}}(\vec{a},\epsilon_1)
     \equiv   - \lim_{ \epsilon_2 \rightarrow 0} p \epsilon_2 \ln Z^{X_{p,q}}_{{\rm full}}
      =  \CW^{\mathbb{C}^2}(\vec{a},\epsilon_1),
         \label{NSlimit}
    \eea
   where we redefined the limit in terms of the equivariant volume of the orbifold.
   This can be derived in the following way.
   In the full partition function \eqref{full} the only part which contributes to this limit 
   is the one with $\ell = L-1$, because the only possibility to get $\frac{1}{\epsilon_{2}}$
   behavior in the exponential is this case 
   as can be seen from the explicit form of $\epsilon_{1,2}^{(\ell)}$ \eqref{several_omega}.
   Then, we see that $Z^{\mathbb{C}^{2}}_{{\rm full}}(\epsilon_{1}^{(L-1)}, \epsilon_{2}^{(L-1)}, \vec{a}^{(L-1)})
   = \exp(\frac{1}{p\epsilon_{2}} F_\Omega(\epsilon_{1}, 0, \vec{a}))$, which leads to \eqref{NSlimit}.

\subsubsection{ALE space}
\label{subsubsec:}
  In this subsection, we consider the expansion \eqref{pippo} in the case of the ALE space.
  The $A_p$ ALE space corresponds to $X_{p,p-1}$.
  In this case, we have 
    \begin{align}
    Z^{\rm ALE}_{{\rm full}}(\epsilon_1, \epsilon_2, \vec{a}, q)
    & \simeq \left( 
           Z^{\mathbb{C}^2}_{{\rm full}} (\epsilon_1, \epsilon_2, \vec{a}, q) 
          \right)^{\frac{1}{p}}
          \exp
          \left[
           \left( - p + \frac{1}{p} \right) F_1(\vec{a},q) 
          \right]
       \cr
           & \times \sum_{\{ \vec{k}^{(\ell)} \}} \xi_{\ell}^{c_{1}^{(\ell)}} \exp
          \left[  \frac{1}{2} \sum_{\ell, m=1}^{p-1}  
            \sum_{\alpha, \beta =1}^{N} 
            k_{\alpha}^{(\ell)} C_{\ell m} 
             \left( \tau_{\rm eff}^{\mathbb{C}^2} \right) ^{\alpha\beta} 
           k_{\beta}^{(m)}
          \right],
    \end{align} 
  since $e_\ell = 2$ for all $\ell$ and $N_{G}=0$.
  
  For illustration, let us consider $A_1$, where the expansion is 
    \bea
    Z^{\rm A_1}_{{\rm full}}(\epsilon_1, \epsilon_2, \vec{a}, q)
    &\simeq& Z^{\mathbb{C}^2}_{{\rm full}} (\epsilon_1, \epsilon_2, \vec{a}, q)^{\frac{1}{2}}
           e^{- 3 F_1(\vec{a},q)/2}
           \sum_{\{ \vec{k} \}} \xi^{c_1} \exp \left[
           \sum_{\alpha, \beta =1}^{N} k_{\alpha}
           \left( \tau_{\rm eff}^{\mathbb{C}^2} \right) ^{\alpha\beta} 
           k_{\beta} \right].
           \label{A1}
    \eea
  Note that in this case $k_\alpha$ is integer or half-integer.
  So far, we did not specify which gauge theory we were considering.
  Now, let us analyze ${\cal N}=2^*$ gauge theory with gauge group $U(2)$.
  We write $F_0$ as
    \bea
    F_0
     =     \pi i \tau_{cl} \sum_{i=1,2} a_i^2 + \tilde{F}_{0},
    \eea
  where $\tilde{F}_{0}$ includes the one-loop and instanton contributions, while the first term is the classical one.
  Since $\tilde{F}_{0}$ only depends on the difference $a_1 - a_2$ in this theory, 
  we can write the coupling constant term as
    \bea
    k_\alpha k_\beta (\tau_{{\rm eff}}^{\mathbb{C}^2})^{\alpha \beta}
     =     k_\alpha k_\beta \partial^\alpha \partial^\beta F_0
     =     \pi i \tau_{{\rm cl}} k_+^2 + \pi i \tau_{{\rm eff}} k_-^2,
    \eea
  where $k_\pm = k_1 \pm k_2$ and we have defined 
  $\tau_{{\rm eff}} = \tau_{{\rm cl}} + \frac{1}{\pi i} \frac{\partial^2}{\partial a_{-}^2}\tilde{F}_{0}$
  with $a_{\pm} = a_{1} \pm a_{2}$.
  Therefore we can rewrite (\ref{A1}) as
    \bea
    Z^{\rm A_1}_{{\rm full}}
    &\simeq&    (Z^{\mathbb{C}^2}_{{\rm full}})^{\frac{1}{2}} e^{- 3 F_1/2}
           \sum_{\{ \vec{k} \}} \xi^{2k_{+}} e^{\pi i \tau_{{\rm cl}} k_+^2} e^{\pi \tau_{{\rm eff}} i k_-^2}
           \nonumber \\
    &=&    (Z^{\mathbb{C}^2}_{{\rm full}})^{\frac{1}{2}} e^{- 3 F_1/2}
           \left( \vartheta_3(q_{{\rm cl}}; \xi^{2}) \vartheta_3(q_{{\rm eff}}; 1)
         + \vartheta_2(q_{{\rm cl}}; \xi^{2}) \vartheta_2(q_{{\rm eff}}; 1) \right),
           \label{A12}
    \eea
  where $q_{{\rm cl}} = e^{\pi i \tau_{{\rm cl}}}$ and $q_{{\rm eff}} = e^{\pi i \tau_{{\rm eff}}}$ and
    \bea
    \vartheta_3 (q; x)
     =     \sum_{n \in \mathbb{Z}} q^{n^2} x^n, ~~~~
    \vartheta_2 (q; x)
     =     \sum_{n \in \mathbb{Z}} q^{(n - 1/2)^2} x^{n-1/2}.
    \eea
  It is interesting to consider the above results for fixed first Chern class.
  For even $c_1$, the first term of the r.h.s.~of (\ref{A12}) contributes as
    \bea
    \frac{Z_{{\rm full}}^{c_1 = even}}{(Z^{\mathbb{C}^2}_{{\rm full}})^{1/2}}
      \simeq q_{{\rm eff}}^{\frac{c_1^2}{4}} e^{-3F_1/2} \vartheta_3(q_{{\rm eff}}).
            \label{thetaoven}
    \eea
  For odd $c_1$, we get
    \bea
    \frac{Z_{{\rm full}}^{c_1 = odd}}{(Z^{\mathbb{C}^2}_{{\rm full}})^{1/2}}
      \simeq q_{{\rm eff}}^{\frac{c_1^2}{4}} e^{-3F_1/2} \vartheta_2(q_{{\rm eff}}).
            \label{thetaodd}
    \eea
  These provide blow-up formulae for {\it singular} points.

\subsubsection{$\CO_{\mathbb{P}^1}(-p)$ space}
  We consider in this subsection the $\CO_{\mathbb{P}^1}(-p)$ space, that is $X_{p,1}$.
  Since in this case there are two patches $L=2$ and $e = p$ as can be seen in Appendix \ref{HJ}, 
  the expansion \eqref{pippo} becomes
    \bea
    & &    Z_{\rm full}^{X_{p,1}}(\epsilon_1, \epsilon_2, \vec{a}, q) 
     \simeq \left( Z_{\rm full}^{\mathbb{C}^2} (\epsilon_1, \epsilon_2, \vec{a}, q) \right)^{\frac{1}{p}}
           \exp \left[ \left( - 2 + \frac{1}{p} \right) F_1(\vec{a},q) + \frac{(p-2)^2}{p} G(\vec{a},q) \right]
           \nonumber \\
    & &    ~~~~~~~
           \times \sum_{\{ \vec{k} \}} \xi^{c_{1}} \exp \left[ 
           \frac{p}{2} \sum_{\alpha, \beta =1}^{N} 
           k_{\alpha} \left( \tau_{\rm eff}^{\mathbb{C}^2} \right) ^{\alpha\beta} k_{\beta}
         + (p-2) \sum_{\alpha = 1}^{N} k_{\alpha} \partial_{\alpha} H \right].
           \label{omp}    
    \eea
  
  Let us specify the above formula for the $U(2)$ gauge theory 
  where the manipulation of the coupling constant term is the same as above.
  We find that in this case the second line in \eqref{omp} becomes
\be
\vartheta_3(q_{{\rm cl}}^{2/p},\xi^2 x_p^+)\vartheta_3(q_{{\rm eff}}^{2/p},x_p^-)
+
\vartheta_2(q_{{\rm cl}}^{2/p},\xi^2 x_p^+)\vartheta_2(q_{{\rm eff}}^{2/p},x_p^-),
\ee
where $x_p^\pm\equiv {\rm exp}\left(\frac{2(p-2)}{p}\partial_{a_\pm}H\right)$.
The above formulas generalize to $\CO_{\mathbb{P}^1}(-p)$ the results in \cite{Nakajima:2003uh}.

\section{Classical, one-loop, and instanton partition functions}
\label{sec:instanton}
  In this section, we separately compute the classical, the one-loop and the instanton contributions 
  for the Nekrasov partition function of the theory on $X_{p,q}$ via consistency of the blow-up formula.
  The classical and the one-loop parts are directly computed by orbifold projection.
  By substituting these parts in the blow-up formula \eqref{full}, we obtain the instanton partition function.
  In this section, we concentrate on the $U(N)$ SYM theory, 
  but all the following results can be easily generalized 
  to the cases with hypermultiplets and to quiver gauge theories.

\subsection{Classical partition function}
\label{subsec:cl}
  As a starting point, let us consider the classical parts. These are given for $\mathbb{C}^2$ and $X_{p,q}$ cases as 
\begin{align}
Z_{\rm cl}^{\mathbb{C}^2} 
(\epsilon_1, \epsilon_2,\vec{a})
   = \exp \left(
      - \frac{\pi i \tau_{\rm cl} \vec{a}^2}{\epsilon_1 \epsilon_2}
     \right) ,
\qquad
Z_{\rm cl}^{X_{p,q}} 
(\epsilon_1, \epsilon_2,\vec{a})
   = \exp \left(
      - \frac{\pi i \tau_{\rm cl} \vec{a}^2}{p \epsilon_1 \epsilon_2}
     \right) .
     \label{treeXpq}
\end{align}
We calculate the contribution from the classical part of the r.h.s.~of (\ref{full}), which is given by
 \begin{align}
 & \exp \left[ 
      - \pi i \tau_{{\rm cl}}\sum_{\ell=0}^{L-1} \sum_{\alpha=1}^N
       \frac{a_{\alpha}^2
        + 2 a_{\alpha} \left( \epsilon_1^{(\ell)} k_{\alpha}^{(\ell+1)} 
        + \epsilon_2^{(\ell)} k_{\alpha}^{(\ell)} \right)
        + \left( \epsilon_1^{(\ell)} k_{\alpha}^{(\ell+1)} 
        + \epsilon_2^{(\ell)} k_{\alpha}^{(\ell)} \right)^2 }%
          {\epsilon_{1}^{(\ell)} \epsilon_{2}^{(\ell)}} 
       \right].
\label{blowexp}
\end{align}
The summation over $\ell$ can be explicitly calculated by using
(\ref{eps-F0}) (\ref{eps-pF0}) and (\ref{eps-tau}).
Then, the classical part of the blow-up formula gives
\begin{align}
\prod_{\ell=0}^{L-1} Z_{\rm cl}^{\mathbb{C}^2}
(\epsilon_1^{(\ell)}, \epsilon_2^{(\ell)},
\vec{a} + \epsilon_1^{(\ell)} \vec{k}^{(\ell+1)} + \epsilon_2^{(\ell)}
\vec{k}^{(\ell)} )
&= Z_{\rm cl}^{X_{p,q}} (\epsilon_1, \epsilon_2,\vec{a})
q^{ \frac{1}{2} \sum_{\ell,m,\alpha}
k_{\alpha}^{(\ell)} C_{\ell m} k_{\alpha}^{(m)} },
\label{tree-blw}
\end{align}
where we have defined $q=e^{2 \pi i \tau}$.

\subsection{One-loop partition function}
\label{subsec:1loop}
The one-loop partition function
on $\mathbb{C}^2$ is given as follows\footnote{Notice that here we need to specify the branch of the $\log$ appearing 
                                               in the perturbative part. 
                                               In order to compare with the results of \cite{Nakajima:2003uh} 
                                               on blow-up formulae, we use their same determination.
                                               This is different from the one chosen in \cite{Alday:2009aq} 
                                               to make comparison with DOZZ three-point functions of Liouville CFT. 
                                               In the zeta-function regularization scheme, 
                                               the two choices are anyway related via analytic continuation 
                                               from $\epsilon_1,\epsilon_2>0$ to $\epsilon_1,\epsilon_2<0$ 
                                               as $\gamma_{\epsilon_1,\epsilon_2}(x)  = -\frac{\pi i}{4} 
                                               + \gamma_{\epsilon_1,\epsilon_2}(-x+\epsilon_1+\epsilon_2)$.
                                               }:
\begin{align}
 Z_{\rm 1-loop}^{\mathbb{C}^2}
  (\epsilon_1, \epsilon_2, \vec{a})
   &= \prod_{\alpha \neq \beta}^N \exp 
      \left[
       - \gamma_{\epsilon_1, \epsilon_2} (a_{\alpha\beta}) 
      \right] ,
      \label{c2vec1loop}
\end{align}
where $a_{\alpha\beta} \equiv a_{\alpha} - a_{\beta}$ and $\gamma_{\epsilon_1, \epsilon_2}$ is 
the logarithm of the Barnes double gamma function:
\bea
\gamma_{\epsilon_1, \epsilon_2} (x) \equiv \log \Gamma(x|\epsilon_1,\epsilon_2).
\eea
For the moment, we assume that $\epsilon_1 > 0$, $\epsilon_2 < 0$.
In this case, the Barnes double gamma function is represented as a regularized 
infinite product as
\begin{align}
\Gamma(x|\epsilon_1,\epsilon_2) = \prod_{m \ge 0, \, n \le -1} (x + m \epsilon_1 + n \epsilon_2).
\label{Gamma_fn}
\end{align}

The one-loop contribution for $X_{p,q}$ is obtained by
projecting \eqref{Gamma_fn} on the part of the spectrum which is invariant under the orbifold action.
We note that the weights of the torus action $(\mathbb{C}^*)^2$ on $\complex^{2}$
are transformed by the $\mathbb{Z}_p$ orbifold action as
\bea
\epsilon_1 \to \epsilon_1 + \frac{2 \pi}{p},
\qquad
\epsilon_2 \to \epsilon_2 + \frac{2 \pi q}{p}.
\eea
The transformation of the weights $a_{\alpha}$ of the Cartan torus $(\mathbb{C}^*)^N$ 
under the orbifold action depends on the magnetic fluxes through the blown-up spheres. 
These are specified by the $u_{(\ell)} \equiv C_{\ell m} k^{(m)}$
and are calculated by imposing that each weight $\vec{a}^{(\ell)}$ of the Cartan torus at the fixed points,
given in (\ref{shift}), is invariant under the orbifold action. 
From \eqref{w-bis} it follows that
\begin{align}
\vec{a} 
&\to \vec{a} 
+ 2 \pi (q_{\ell} \vec{k}^{(\ell+1)} - q_{\ell+1} \vec{k}^{(\ell)})
\qquad ({\rm mod} \,\, 2\pi).
\label{a_orb}
\end{align}
Since 
\bea
2 \pi (q_{\ell} \vec{k}^{(\ell+1)} - q_{\ell+1} \vec{k}^{(\ell)})
- 2 \pi (q_{\ell-1} \vec{k}^{(\ell)} - q_{\ell} \vec{k}^{(\ell-1)})
= 2\pi q_{\ell} \vec{u}_{(\ell)}
\eea
from the explicit calculation,
we see that (\ref{a_orb}) gives the consistent expression for arbitrary $\ell$.
In the following, we choose $\ell=0$, which simplifies (\ref{a_orb}) as
\begin{align}
a_{\alpha} 
&\to a_{\alpha} 
- 2 \pi k^{(1)}_\alpha
\qquad ({\rm mod} \,\, 2\pi).
\end{align}
Note that fixing $k_{\alpha}^{(1)}$ determines the holonomy of the gauge field at infinity.

Therefore, the one-loop factor for the theory on $X_{p,q}$ 
depends on $k_{\alpha\beta}^{(1)} \equiv k_{\alpha}^{(1)} - k_{\beta}^{(1)}$ 
mod $\mathbb{Z}$, which divides the partition function into $p^{N}$ sectors.
These are obtained by replacing $\gamma_{\epsilon_1, \epsilon_2} (a_{\alpha\beta})$ 
in $Z^{\mathbb{C}^2}_{\rm 1-loop}$ by 
\bea
\tilde{\gamma}_{\epsilon_1, \epsilon_2} (a_{\alpha\beta},pk^{(1)}_{\alpha\beta})
 &\equiv& \sum_{{m \ge 0, \, n \le -1} \atop {m+nq=pk^{(1)}_{\alpha\beta}} \,\, {\rm mod} \,\, p} 
          \log ( a_{\alpha\beta} + m\epsilon_1+n\epsilon_2)
 \nonumber \\
 &=&      \sum_{{qn \le pm, \, n \le -1} \atop {(m,n)=(k^{(1)}_{\alpha\beta},0) \,\, {\rm mod} \,\, \mathbb{Z}^2 } } 
          \log \left( a_{\alpha\beta} + m \epsilon_1^{(0)} + n \epsilon_2^{(0)} \right) ,
 \label{gamma_t}
\eea
where $\epsilon_1^{(0)}=p\epsilon_1$, $\epsilon_2^{(0)}= - q\epsilon_1 + \epsilon_2$.
That is,
\bea
 Z^{X_{p,q}}_{\rm 1-loop} (\vec{a}, \epsilon_1, \epsilon_2, p\vec{k}^{(1)})
= \prod_{\alpha \neq \beta} \exp \left( 
   - \tilde{\gamma}_{\epsilon_1, \epsilon_2} 
   (a_{\alpha\beta}, pk^{(1)}_{\alpha\beta})
  \right).
\eea
We note that $Z^{X_{p,q}}_{\rm 1-loop}$ depends on $p\vec{k}^{(1)}$ by mod $p$.

We calculate the contribution from the one-loop part of the r.h.s.~of (\ref{full}).
Note that if $\epsilon_1>0, \epsilon_2<0$, then $\epsilon^{(\ell)}_1>0, \epsilon^{(\ell)}_2<0$.
This follows from the convexity of the dual fan, which indicates $p_{\ell} / q_{\ell} > p / q$ for $2 \le \ell \le L$.
Therefore, we can use the expression (\ref{Gamma_fn}) for each one-loop factor.

The key identity to construct the blow-up formula at one-loop is 
\bea
\sum_{\ell=0}^{L-1} 
  \gamma_{\epsilon_1^{(\ell)}, \epsilon_2^{(\ell)}} 
  (a_{\alpha\beta}^{(\ell)} ) 
 &=& \tilde{\gamma}_{\epsilon_1, \epsilon_2} (a_{\alpha\beta},pk^{(1)}_{\alpha\beta} )
      + \sum_{\ell=0}^{L-1} 
     f_{p,q}^{(\ell)}(a_{\alpha\beta}^{(\ell)}, \epsilon_1^{(\ell)}, \epsilon_2^{(\ell)}, k^{(\ell)}_{\alpha\beta}, k^{(\ell+1)}_{\alpha\beta}) ,
\label{keyId}
\eea
where $\tilde{\gamma}_{\epsilon_1, \epsilon_2} (a_{\alpha\beta},k^{(1)}_{\alpha\beta})$ corresponds to
the one-loop factor for $X_{p,q}$ while $f_{p,q}^{(\ell)}$ is the finite sum
\bea
 f_{p,q}^{(\ell)}(a, e_1, e_2, \mu, \nu) 
 \equiv \left\{
 \begin{array}{cll}
 \displaystyle \sum_{{m \ge 0, n \le -1}\atop {p_{\ell} (\nu+m) \le p_{\ell+1} (\mu+n)} } 
 \log \left( 
   a + m e_1 + n e_2
 \right)
 & \qquad & p_{\ell} \nu < p_{\ell+1} \mu
 \\
 0 & \qquad & p_{\ell} \nu = p_{\ell+1} \mu
 \\  \\
 \displaystyle \sum_{{m \le -1, n \ge 0}\atop 
     {p_{\ell} (\nu+m) > p_{\ell+1} (\mu+n)} } 
 \log \left( 
   a + m e_1 + n e_2
 \right)
 & \qquad & p_{\ell} \nu > p_{\ell+1} \mu
 \end{array}
 \right.
\label{def_f}
\eea
Note that 
the condition defining the $f_{p,q}^{(\ell)}$ function weights differently the contributions
accordingly to the sign of $-p_{\ell} k^{(\ell+1)} + p_{\ell+1} k^{(\ell)}$,
that is the sign of the coefficient of $\epsilon_2^{(0)}$ in 
$\epsilon_1^{(\ell)} k_{\alpha}^{(\ell+1)} + \epsilon_2^{(\ell)} k_{\alpha}^{(\ell)} $
expanded using \eqref{e1le2l}.
By analytic continuation \eqref{keyId} and \eqref{def_f} are valid for any complex values of $\epsilon_1$ and $\epsilon_2$.

\begin{figure}[t]
\centering
\includegraphics[width=8cm]{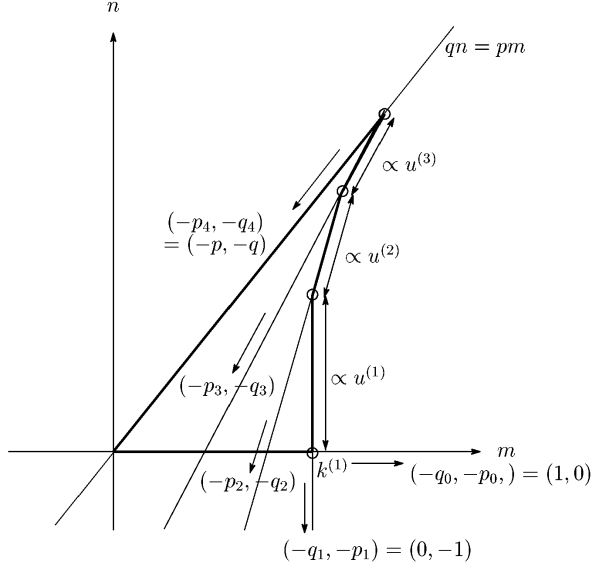}
\vspace{5mm}
\caption{Pictorial rendering of the identity (\ref{keyId}) for $L=4$}
\label{fig:Lfactor}
\end{figure}

The identity (\ref{keyId}) can be understood pictorially when $u^{(\ell)} > 0$ for $\forall \ell$,
as depicted in Figure \ref{fig:Lfactor} for $L=4$.
In this figure, each points $(m,n)$ corresponds to the factor
\bea
\log (a_{\alpha\beta} + m\epsilon_1^{(0)} + n\epsilon_2^{(0)}).
\eea
Since 
\bea
\epsilon_1^{(\ell)} 
  =    - q_{\ell} \epsilon_1^{(0)} - p_{\ell}  \epsilon_2^{(0)},
\qquad
\epsilon_2^{(\ell)} 
  =    q_{\ell+1} \epsilon_1^{(0)} + p_{\ell+1}  \epsilon_2^{(0)},
\label{e1le2l}
\eea
each of the summand in the left hand in (\ref{keyId}) adds lattice points 
between the lines generated by the vectors $(-q_{\ell}, -p_{\ell})$ and $(-q_{\ell+1}, -p_{\ell+1})$ 
starting from the points corresponding to $a^{(\ell)}_{\alpha\beta}$ and $a^{(\ell+1)}_{\alpha\beta}$ respectively.
The first term in the r.h.s.~corresponds to the points in the region $qn \le pm, \, n \le -1$ by definition.
The remaining term corresponds to the points 
in the region surrounded by the bold lines in Figure \ref{fig:Lfactor}.
It is remarkable that the boundary of this region 
consists of the lines $n=0$, $qn=pm$, and the generating vectors of the dual fan,
each of which length is proportional to the magnetic fluxes 
$u_{\alpha\beta}^{(\ell)} \equiv u_{\alpha}^{(\ell)} - u_{\beta}^{(\ell)} $.
We also note that the region 
is naturally divided into $L-1$ regions,
which is interpreted as the contribution from each blown-up sphere.
From (\ref{aellalpha}) and (\ref{e1le2l}), 
we see that the point corresponding to $a^{(\ell)}_{\alpha\beta}$ 
satisfies the condition
\bea
(m,n) =(k^{(1)}_{\alpha\beta},0) \,\, {\rm mod} \,\, \mathbb{Z}^2,
\label{condMN}
\eea
which is part of the condition for the summation in (\ref{gamma_t}).
Moreover, the basis $\epsilon_1^{(\ell)}$ and $\epsilon_2^{(\ell)}$ generates
the lattice points exactly the same as those generated by $\epsilon_1^{(0)}$ and $\epsilon_2^{(0)}$ because $p_{\ell} q_{\ell+1} - q_{\ell} p_{\ell+1} = 1$.
Therefore, each term in (\ref{keyId}) adds the points
satisfying the condition (\ref{condMN}) inside its relative region as previously stated and this
explains the identity (\ref{keyId}).

By using the identity (\ref{keyId}), we find that the blow-up formula for the one-loop part is given by
\begin{align}
&\prod_{\ell=0}^{L-1} Z_{\rm 1-loop}^{\mathbb{C}^2}
\left( \epsilon_1^{(\ell)}, \epsilon_2^{(\ell)},
\vec{a}^{(\ell)} 
\right)
= \ell _{\rm vector} (\epsilon_1, \epsilon_2,\{\vec{k}^{(\ell)} \})
 Z^{X_{p,q}}_{\rm 1-loop} (\epsilon_1, \epsilon_2, \vec{a}, p\vec{k}^{(1)}),
\label{1lp-blw}
\end{align}
where
\begin{align}
& \ell _{\rm vector} ( \epsilon_1, \epsilon_2,\{\vec{k}^{(\ell)} \} )
  = \prod_{\ell=0}^{L-1} \prod_{\alpha \neq \beta}
    \exp \left( 
     - f_{p,q}^{(\ell)}(a^{(\ell)}_{\alpha\beta}, \epsilon_1^{(\ell)}, 
     \epsilon_2^{(\ell)}, k^{(\ell)}_{\alpha\beta}, k^{(\ell+1)}_{\alpha\beta})
    \right).
\label{ellfact}
\end{align}

\subsection{Instanton partition function}

The instanton partition function is obtained by combining the 
full blow-up formula (\ref{full}) with 
the results in section \ref{subsec:cl} and \ref{subsec:1loop}.
Since the one-loop factor depends on $\vec{k}^{(1)}$
mod $\mathbb{Z}$, the full partition function 
is written by the sum of these sectors,
\bea
Z_{\rm full}^{X_{p,q}} (\epsilon_1,\epsilon_2,\vec{a}, \{ \xi_{\ell} \})
 = Z_{\rm cl}^{X_{p,q}}(\epsilon_1,\epsilon_2,\vec{a})
    \displaystyle \sum_{I_1=0}^{p-1} \cdots \sum_{I_N=0}^{p-1}
     &&Z_{\rm 1-loop}^{X_{p,q}} 
     \left(
     \epsilon_1,\epsilon_2, \vec{a}, \vec{I}
     \right)
     Z_{\rm inst}^{X_{p,q}} 
     \left(
     \epsilon_1,\epsilon_2, \vec{a}, \vec{I}, \{ \xi_{\ell} \}
     \right)
  \nonumber \\
\eea
where $\vec{I}=p\vec{k}^{(1)}$ mod $p$ parametrizes the holonomy class of the gauge field in the Cartan torus.
Finally, by using the classical (\ref{tree-blw}), the one-loop (\ref{1lp-blw}), 
and the full blow-up formula (\ref{full}), we obtain the instanton partition function as
    \begin{align}
 Z_{\rm inst}^{X_{p,q}} 
 \left(
 \epsilon_1,\epsilon_2, \vec{a}, \vec{I}, \{ \xi_{\ell} \}
 \right) 
  = & \sum_{ { \{ \vec{u}_{(\ell)} \} \in \mathbb{Z}^{N(L-1)}}
       \atop{ p\vec{k}^{(1)}=\vec{I} \,\, {\rm mod} \, p}}
       q^{ \frac{1}{2} \sum_{\ell,m}
        \vec{k}^{(\ell)} \cdot C_{\ell m} \vec{k}^{(m)}}
       \times \ell_{\rm vector} ( \epsilon_1, \epsilon_2,\{\vec{k}^{(\ell)} \} )
     \cr
       & \qquad \qquad \times
       \prod_{\ell=0}^{L-1} Z_{\rm inst}^{\complex^2}
       \left( \epsilon_1^{(\ell)}, \epsilon_2^{(\ell)},
       \vec{a}^{(\ell)} \right)
     \times
     \prod_{\ell=1}^{L-1} \xi_{\ell}^{\sum_{\alpha=1}^N u_{\alpha (\ell)}},
 \label{blw-inst}
 \end{align}
where $u_{\alpha (\ell)} = C_{\ell m} k_{\alpha}^{(m)}$.

\subsection{ALE space}

In this subsection we explicitly consider the above formulae for the ALE case, namely $q = p-1$.
In this case, $L=p$, and $p_{\ell}=\ell$.
The classical partition function is given by (\ref{treeXpq}).
The one-loop partition function is given  as
\begin{align}
 Z^{\rm ALE}_{\rm 1-loop} (\epsilon_1, \epsilon_2, \vec{a}, \{ \vec{k}^{(\ell)} \} )
= & \prod_{\ell=0}^{p-1} \prod_{\alpha \neq \beta}
g^{(\ell)} (a^{(\ell)}_{\alpha\beta}, \epsilon_1^{(\ell)}, 
     \epsilon_2^{(\ell)}, k^{(\ell)}_{\alpha\beta}, k^{(\ell+1)}_{\alpha\beta})
\prod_{\ell=0}^{p-1} Z_{\rm 1-loop}^{\mathbb{C}^2}
\left( \epsilon_1^{(\ell)}, \epsilon_2^{(\ell)},
\vec{a}^{(\ell)} 
\right)
\end{align}
with
\bea
 g^{(\ell)} (a, e_1, e_2, \mu, \nu) 
 \equiv \left\{
 \begin{array}{cll}
 \displaystyle \prod_{{m \ge 0, n \le -1}\atop \ell (\nu+m) \le (\ell+1) (\mu+n)}
 \left( 
   a + m e_1 + n e_2
 \right)
 & \qquad & \ell \nu < (\ell+1)\mu 
 \\
 1 & \qquad & \ell \nu = (\ell+1) \mu
 \\  \\
 \displaystyle \prod_{{m \le -1, n \ge 0}\atop 
     \ell (\nu+m) > (\ell+1) (\mu+n)}
 \left( 
   a + m e_1 + n e_2
 \right)
 & \qquad & \ell \nu > (\ell+1) \mu 
 \end{array}
 \right. .
\eea
Here $a_{\alpha}^{(\ell)}$ is defined in (\ref{shift}) and 
$\epsilon_1^{(\ell)}$, $\epsilon_2^{(\ell)}$ are given in (\ref{w-bis}).
Note that the dependence on $\{ \vec{k}^{(\ell)} \}$ actually reduces to the dependence on $p\vec{k}^{(1)}$ mod $p$.
The instanton partition function is given by 
 \begin{eqnarray}
 Z_{\rm inst}^{\rm ALE} 
 \left(
 \epsilon_1,\epsilon_2, \vec{a}, \vec{I}, \{ \xi_{\ell} \}
 \right) 
  &&= \sum_{ { \{ \vec{u}_{(\ell)} \} \in \mathbb{Z}^{N(p-1)}}
       \atop{ p\vec{k}^{(1)}=\vec{I} \,\, {\rm mod} \, p}}
       q^{ \frac{1}{2} \sum_{\ell,m}
        \vec{k}^{(\ell)} \cdot C_{\ell m} \vec{k}^{(m)}}
        \prod_{\ell=0}^{p-1} 
        g^{(\ell)} (a^{(\ell)}_{\alpha\beta}, \epsilon_1^{(\ell)}, 
        \epsilon_2^{(\ell)}, k^{(\ell)}_{\alpha\beta}, k^{(\ell+1)}_{\alpha\beta})
        {}^{-1}
     \nonumber \\
      & & \qquad \qquad \times
       \prod_{\ell=0}^{p-1} Z_{\rm inst}^{\complex^2}
       \left( \epsilon_1^{(\ell)}, \epsilon_2^{(\ell)},
       \vec{a}^{(\ell)} \right)
     \prod_{\ell=1}^{p-1} \xi_{\ell}^{\sum_{\alpha=1}^N u_{\alpha (\ell)}}.
\label{aleinst} 
\end{eqnarray}
We note that $C$ is the $A_{p-1}$ Cartan matrix and $\vec{I}$ specifies the twisted sector:
\bea
\vec{a} \to \vec{a} - \frac{2 \pi \vec{I}}{p}.
\eea
In appendix \ref{sec:explicit} we list the explicit first orders of the instanton partition function 
of $U(2)$ SYM theory on the $A_2$ ALE space.


\section{M-theory on toric singularities and blow-up formulae}
\label{sec:M}

In this section we discuss how the blow-up formulae in the $\epsilon_1,\epsilon_2\to 0$ limit 
of section \ref{sec:gauge} can be derived from M-theory considerations. 
As we will shortly explain, these can be obtained by considering the classical partition function
of a single M5-brane on a suitable product geometry. 
The $\epsilon$-corrections should be calculable from the quantum contribution
to the M5-partition function at least in the $\epsilon_1 + \epsilon_2 =0$ case
via a generalization of \cite{Dijkgraaf:2002ac}, but we will not discuss this issue in the present paper.
  
Let us recall that the low-energy effective theory of $\mathcal{N}=2$ four-dimensional gauge theory
is described by a single space-time filling M5-brane wrapping the Seiberg-Witten curve \cite{Witten:1997sc},
so that the appropriate six-dimensional manifold is the direct product $M_6=\Sigma\times X_{p,q}$.
We will show that the M5-brane partition function can be computed in this setting. Let us first quickly
review the case of compact six-manifolds and then specify the modifications relevant to the non-compact case.
  
The world-volume theory of the M5-brane is described by an anti-self-dual tensor multiplet $T$ minimally coupled
with a three-form gauge field $C_3$ by a term $\int_{M_6} T\wedge {C}_3$.
Let us consider a symplectic basis $E_A = (e^a,\tilde e_b)$ for the middle cohomology $H^3(M_6,\zet)$ of the six-manifold $M_6$
\be
\int_{M_6} E_A\wedge E_B = J_{AB},
\ee
namely
\be
\int_{M_6} e^a\wedge e^b = 0, \ \ \int_{M_6} \tilde e_a\wedge \tilde e_b = 0, \ \ 
\int_{M_6} e^a\wedge \tilde e_b = \delta^a_{b},
\label{simpl}
\ee
where $J_{AB} = \1 \otimes${\tiny $\begin{pmatrix} 0 & 1\\ -1 & 0 \end{pmatrix}$}.
Then the period matrix $Z=Z^{(1)}+iZ^{(2)}$ is defined by expanding 
\be
\tilde e_a = Z^{(1)}_{ab} e^b + Z^{(2)}_{ab} \star e^b.
\label{period}
\ee 
The simplest way to get the M5-brane classical contribution is by computing the full partition function
of an abelian two-form potential $T=dB$ in six dimensions. This holomorphically factorizes in chiral times anti-chiral blocks.
The M5-brane partition function is extracted as one of such chiral blocks, in the very same way as chiral partition functions
are obtained in two-dimensions \cite{Verlinde:1995mz,Witten:1996hc,Henningson:1999dm,Bonelli:2001pu,Bonelli:2001jf}. 
The explicit form of this partition function is given  
in terms of a theta function $\Theta(\overline Z)$ with suitable characteristics and arguments. 
The arguments of the theta function
are related to the period matrix of the six-manifold and the background value of the $C$-field.
 
This approach can be extended also to non-compact six-manifolds. In this case, one should restrict to the
$L^2$ sector of the middle cohomology in order to count finite action configurations only.
In particular, let us consider $M_6=\Sigma\times X_{p,q}$. In this case, the
K\"unneth decomposition of the middle cohomology reads
\be
H^3(\Sigma\times X_{p,q}, \zet )=H^1(\Sigma,\zet)\otimes H^2(X_{p,q},\zet).
\label{kunneth}
\ee
Let us choose a polarization given by a symplectic basis
\bea
e^{\alpha (n)} &=& [b^\alpha] \wedge \omega^n,
 \nonumber\\
\tilde e_{\alpha (m)} &=& - [a_\alpha]\wedge {\cal I}_{m\ell} \omega^\ell,
\label{basis}
\eea
where
$[a_\alpha]$ and $[b^\alpha]$ are the Poincar\'e duals of the $a$ and $b$ cycles on $\Sigma$
and ${\cal I}$ satisfies 
\be
{\cal I}_{m\ell}\int_{X_{p,q}}\omega^{\ell}\wedge \omega^n=\delta_m^n  \ \ . 
\ee
The $L^2$ sector of the middle cohomology of $X_{p,q}$ coincides with the anti-selfdual 
part of $H^2(X_{p,q},\zet)$ \cite{Barth}.
Therefore, in the subspace which is relevant to our computation,
$\star \omega^m= -\omega^m$ and, by choosing the normalization  
$\int_{\Delta_\ell}\omega^n=\delta_\ell^n$, we get that
${\cal I}=-C$ is 
nothing but the intersection matrix of $X_{p,q}$ homology cycles, see Appendix \ref{HJ}
for details. In our notation then $b^-\left(X_{p,q}\right)=L-1$.

To compute the period matrix, one can use the standard relation 
$$
[a_\alpha]=-\tau^{(1)}_{\alpha\beta}[b^\beta]-\tau^{(2)}_{\alpha\beta}\star[b^\beta]
$$
which holds on a generic Riemann surface \cite{FK}, where $\tau=\tau^{(1)}+i\tau^{(2)}$
is its period matrix.
Indeed one gets
$$
\tilde e_{\alpha (m)}=[a_\alpha] \wedge C_{mn}\omega^n=\left\{
-\tau^{(1)}_{\alpha\beta}[b^\beta]-\Omega^{(2)}_{\alpha\beta}\star[b^\beta]
\right\} \wedge
C_{mn}\omega^n=
\left(-\tau^{(1)}_{\alpha\beta}C_{mn}\right)e^{\beta (n)}+\left(\tau^{(2)}_{\alpha\beta}C_{mn}\right)\star e^{\beta (n)}
$$
so that the complete period matrix relevant to finite action configurations is 
$$
Z=-C\otimes\bar\tau.
$$
Another subtle feature arise in the non-compact case when considering the 
possible allowed three- form fluxes. 
Let us show that the most natural condition in M-theory leads to the correct lattice of four-dimensional gauge theory.
First of all we expand the three-form $T\in H^3(\Sigma\times X_{p,q})$ as
\be
T = \hat k_{\alpha (m)} e^{\alpha (m)} + \hat h^{\alpha (m)} \tilde e_{\alpha (m)}
\label{su}
\ee 
and compute its periods on generic compact 3-cycles as
\be
\int_{\Gamma^a} T = \hat k_{\alpha (m)}\int_{\gamma^i}[b^\alpha]\int_{\Delta_n} \omega^m 
- \hat h^{\alpha (m)} \int_{\gamma^i} [a_\alpha] \int_{\Delta_n}\omega^l C_{lm} 
=\left\{ \begin{array}{ll}
         \hat k_{\alpha (m)} (C^{-1})^{mn} & \ \ \mbox{if $i = \alpha $};\\
         \hat h^{\alpha (m)} & \ \ \mbox{if $i = g + \alpha $}.\end{array} \right.
\label{zio}
\ee
Here $\left\{\Gamma^a\right\}=\left\{\Delta_n\times\gamma^i\right\}$, where
$\left\{\gamma^i\right\}=\left\{a_\alpha,b^\alpha\right\}$ is our reference basis in $H_1(\Sigma,\zet)$
and $\left\{\Delta_n\right\}$ are the blown-up spheres in $X_{p,q}$.

The lattice of the relevant theta function is obtained by dualizing the lattice of the $\{e^{a}\}$ sector \cite{Henningson:1999dm},
namely the first line in the r.h.s.~of (\ref{zio}).
Being this the lattice of fundamental M-theory charges we assume that it is the integer one, namely 
$\left(C^{-1}\otimes{\bf 1}\right) \hat k\in \zet^{g(L-1)}$.
Then its dual reads $\left(C\otimes{\bf 1}\right) k\in \zet^{g(L-1)}$ so that finally
\be
k 
\in (C^{-1}\otimes {\bf 1})\zet^{g(L-1)}
\ee
whose four-dimensional factor describes the magnetic fluxes of the gauge field on $X_{p,q}$. 
Therefore, the relevant generalized $\Theta$-function contribution is given by the
anti-holomorphic block which reads
\be
\Theta
(C\otimes\tau|Q)=\sum_{k \in (C^{-1}\otimes {\bf 1})\zet^{g(L-1)}}
e^{\pi i \left[ k (C\otimes\tau) k + 2 Q
(C\otimes {\bf 1})
k \right]}
\label{theta}
\ee
where $Q$ is a background for the three-form $C$-field which couples to the non-trivial $c_1$ of the
four-dimensional gauge bundle. Indeed from the minimal coupling term we get 
\be
\int_{M_6} C_3\wedge T
=
\int_{\Sigma\times X_{p,q}} C_3\wedge\left(\hat k_{\alpha (m)} e^{\alpha (m)} + \hat h^{\alpha (m)} \tilde e_{\alpha (m)}\right)
=\sum_{\alpha (m)} \left(\hat Q^{\alpha (m)} \hat k_{\alpha (m)}+\tilde Q_{\alpha (m)} \hat h^{\alpha (m)}\right)
\ee
where $\hat Q^a$ and $\tilde Q_a$ are the components of the background $C$-field.
After Poisson resummation, one gets the linear term $kQ$ in (\ref{theta}) which 
reproduces the term $\left(\xi_{m}\right)^{ C_{lm} k_{\alpha}^{(m)}}$ in the gauge theory $\theta$-function.
The complete background field is actually 
$Q^{\alpha (m)}=\frac{1}{2\pi i}\ln \xi_{m}+Q^{\alpha (m)}_{grav}$ whose first term 
describes, after the reduction, the coupling of an external 
source to the first Chern class of the gauge bundle.
The second term $Q_{grav}$ has to be set by considerations of anomaly cancellation conditions.
In order to address this issue, one should compute the full quantum mechanical contribution to the 
partition function in the product geometry.
The requirement of modular invariance then would correspond to the anomaly cancellation conditions in the 
string theory engineering.
In general there are pure gravitational and mixed anomalies to cure.
Fixing the pure gravitational part means fixing the overall normalization of the M5-brane partition function, 
namely the coefficients of the $F_1$ and $G$ terms in (\ref{pippo}), while fixing 
the mixed anomalies corresponds to fix the coefficients of the linear term in the $k_\alpha^{(\ell)}$.
This corresponds to fix the gravitational part of the background $C$-field
as $Q^{\alpha (m)}_{grav}= \partial_\alpha H (e_\ell-2)$.
Although we don't give an M-theoretic detailed account for this assignment, let us notice that 
it measures an anomaly induced the lack of the four manifold from being hyperkahler, that is from 
admitting a reduced holonomy which would make the topologically twisted and the untwisted ${\cal N}=2$
gauge theories equivalent.

It is then easy to recognize that, after the above background value is selected, (\ref{theta}) 
coincides with the ($SU(N)$ part of the)
$\vartheta$-function
arising in the blow-up formulae of the previous sections (\ref{pippo}).

\section{Two-dimensional CFT counterpart}
\label{sec:CFT}
  In this section, we consider the relation between the gauge theories on toric singularities and two-dimensional CFTs.
  The M-theory argument in the previous section gives a hint to this relation: 
  one side is the four-dimensional gauge theory on $X_{p,q}$ 
  obtained by compactifying M5-branes on a punctured Riemann surface $\CC$.
  The other side is the two-dimensional  effective theory on  $\CC$ 
  obtained after integration over the four-dimensional modes of the M-theory.
  As it is well-known, the two-dimensional theory in the case of $\mathbb{R}^{4}_{\epsilon_{1}, \epsilon_{2}}$ 
  turns out to be the $W_{N}$ CFT \cite{Alday:2009aq, Wyllard:2009hg} with an extra $U(1)$ free boson factor.
  The partition function of the gauge theory can be related with the conformal block $\CB_{W}$ of $W_{N}$-algebra, 
  under a proper identification of the parameters, 
    \bea
    Z^{\mathbb{C}^{2}}
     =     Z_{{\rm U(1)}} \CB_{W},
    \eea
  where $\CB_{W}$ is defined in a specific marking of $\CC$
  which in turn corresponds to a particular weak coupling description of the gauge theory.
  Non-trivial four-dimensional geometry should change the two-dimensional CFT algebra to ${\cal A}_N(X_{p,q})$.
  Indeed, it was found and checked 
  in \cite{Belavin:2011pp, Nishioka:2011jk, Bonelli:2011jx, Belavin:2011tb, Bonelli:2011kv, 
  Wyllard:2011mn, Ito:2011mw, Alfimov:2011ju, Belavin:2011sw}
  that the U($N$) gauge theory on the $A_{p-1}$ ALE space is related 
  to the product of the Heisenberg, affine $\hat{su}(p)_{N}$ and $\mathbb{Z}_{p}$ para-$W_{N}$ algebras.
  In this section, we study this relation by using the gauge theory partition function computed 
  in section \ref{sec:gauge} and the M-theory considerations presented in section \ref{sec:M}, 
  and investigate their possible implications on the mysterious two-dimensional side.
  
  The blowup formula \eqref{full} directly imply that the fixed points of the moduli space of instantons 
  on $X_{p,q}$ provide a basis for the representation of $L$ copies of Heisenberg plus $W_N$ algebrae 
  with suitable central charges related to the $(\epsilon_1^{(\ell)}, \epsilon_2^{(\ell)})$ weights 
  under the $(\mathbb{C}^*)^2$-action
    \be
    {\cal A}_N(X_{p,q})\equiv \oplus_{\ell=0}^{L-1} (\mathcal{H} \oplus \, ^\ell W_N  )
    \label{copies}
    \ee
  where the central charge of $^\ell W_N$ is $c_\ell = (N-1)\left(1+ Q_\ell^2 N(N+1)\right) $, 
  $Q_\ell^2 = \frac{(\epsilon_1^{(\ell)})^2+(\epsilon_2^{(\ell)})^2}{\epsilon_1^{(\ell)}\epsilon_2^{(\ell)}} + 2$.
  As we will see in a moment, the overall central charge of \eqref{copies} coincides 
  with the central charge of the two-dimensional CFT that can be computed from M-theory compactification.  

  The result \eqref{copies} suggest that in the ALE case $q=p-1$ 
  the conformal blocks of $\zet_p$ parafermionic $W_N$ algebra plus $p$ copies of Heisenberg algebrae can be computed 
  as $p$-nested conformal blocks of the algebra \eqref{copies} in terms of the blowup formula \eqref{aleinst}. 
  This has been shown in \cite{Bonelli:2011jx, Bonelli:2011kv, Belavin:2011sw} in the $p=2$ case, 
  corresponding to the ${\cal N}=1$ super Virasoro algebra whose conformal blocks are known. 
  The conformal blocks for the general case are instead unknown and the blowup formula \eqref{aleinst} provides 
  a natural candidate for them.

\subsection{$U(1)$ partition function and Frenkel-Kac construction}
\label{subsec:Frenkel}
From the mathematical viewpoint, the above correspondence for the ALE case should be viewed 
as a generalization to higher rank cases of the Frenkel-Kac construction discussed 
in \cite{FrenkelKac, nagao} for Hilbert schemes, {\it i.e.} $U(1)$ gauge group.
In this subsection, we consider the case of $U(1)$ gauge theory 
and the relation with the Frenkel-Kac construction more explicitly.

For the abelian case, the instanton partition function (\ref{aleinst}) reduces to
\begin{align}
Z_{\rm inst}^{\rm ALE} 
= \sum_{\{ u_{(\ell)} \}} q^{\frac{1}{2} \sum_{\ell,m} u_{(\ell)}(C^{-1})^{\ell m}u_{(m)}}
\prod_{\ell=0}^{p-1} 
Z_{\rm inst}^{\mathbb{C}^2} (\epsilon^{(\ell)}_1, \epsilon^{(\ell)}_2, a^{(\ell)})
\prod_{\ell=1}^p \xi_{\ell}^{u_{(\ell)}}
\label{abel_inst}
\end{align}
since $k_{\alpha\beta}=0$, and thus $g^{(\ell)}=1$. 
This is also the case when we consider ${\cal N}=2^*$ theory,
where $Z_{\rm inst}^{\mathbb{C}^2}$ is replaced by 
the Nekrasov instanton partition function of the ${\cal N}=2^*$ theory on $\mathbb{C}^2$.
In the ${\cal N}=4$ limit, it further reduces to
\begin{align}
Z_{\rm inst}^{\rm ALE} 
= \eta(\tau)^{-p} \sum_{ \{ u_{(\ell)} \} } q^{\frac{1}{2} \sum_{\ell,m} u_{(\ell)}(C^{-1})^{\ell m}u_{(m)}}
\prod_{\ell=1}^p \xi_{\ell}^{u_{(\ell)}}.
\label{N=4}
\end{align}
which can also be written as
\begin{align}
Z_{\rm inst}^{\rm ALE}  = \eta^{-1}(\tau) \sum_{\Lambda} \chi_{\Lambda} (\tau, \xi),
\label{N=4'}
\end{align}
where $\chi_{\Lambda}$ is the character of the representations $\Lambda$ of the affine $\hat{su}(p)_1$ 
algebra, as found in \cite{Bianchi:1996zj}.

On one hand, one can see that \eqref{abel_inst} or \eqref{N=4} are related 
to the direct sum of $p$ copies of the Heisenberg algebra in the two-dimensional CFT side. 
On the other hand, \eqref{N=4'} denotes the relation with the algebra $\mathcal{H} \oplus \hat{su}(p)_1$
as in \cite{Belavin:2011sw}.
These two different sets of algebras in the two-dimensional CFT side can be understood 
as the Frenkel-Kac construction \cite{FrenkelKac}: actually,
by this construction, $\hat{su}(p)_1$ is expressed as the Heisenberg algebra
of the type $A_{p-1}$:
\begin{equation}
[a_{i (\ell)}, a_{j (m)}] = \delta_{i+j, 0} C_{\ell m},
\end{equation}
where $C_{\ell m}$ is the Cartan matrix for $A_{p-1}$. 
The corresponding CFT is the theory of free bosons, whose Hamiltonian is given by
\begin{align}
H = \sum_{\ell,m} \left(
\frac{1}{2} a_{0 (\ell)} (C^{-1})^{\ell m} a_{0 (m)}
+ \sum_{i=1}^{\infty}  a_{-i (\ell)} (C^{-1})^{\ell m} a_{i (m)}
\right).
\end{align}
From the viewpoint of M-theory, the two-dimensional CFT is obtained by 
reducing the theory of self dual three-form on $T^2 \times \mathbb{C}^2/Z_p$ onto $T^2$.
Therefore, the target space of the two-dimensional CFT is expected to be
the integrable second cohomology of the ALE space modulo integral elements and thus,
the momentum lattice is given by that of $A_{p-1}$.
The partition function on a torus with the insertion of the fugacity $z_{(\ell)}$
for this theory,
which corresponds to the ${\cal N}=4$ theory in the gauge theory side,
is given by
\begin{align}
{\rm Tr}  \langle q^{H}
e^{\sum_{\ell} (C^{-1} z)^{(\ell)} a_{0 (\ell)} 
} \rangle 
= \eta(\tau)^{-p+1} \sum_{ \{ u_{(\ell)} \} } q^{\frac{1}{2} \sum_{\ell,m} u_{(\ell)}(C^{-1})^{\ell m}u_{(m)}}
\prod_{\ell=1}^{p-1} \xi_{(\ell)}^{u_{(\ell)}},
\label{2DCFT_part}
\end{align}
where the integers $u_{(\ell)}$ corresponds to the momenta.
By multiplying by the partition function $\eta^{-1}(\tau)$ of one more free 
boson\footnote{In the M-theory derivation one obtains this as the overall 
contribution of the remaining part of the reduced selfdual multiplet.},
which corresponds to the overall $U(1)$ factor in the gauge theory side,
it exactly reproduces (\ref{N=4}).

We expect that the factorization property in (\ref{2DCFT_part}) holds 
even when we add one-point insertion with finite conformal dimension. 
In this case, $\eta^{-1}$ function is replaced by the 
one-point function of the free boson on torus,  
which is known to coincide with 
the Nekrasov partition function of the abelian ${\cal N}=2^*$ theory
on $\mathbb{C}^2$ \cite{Nekrasov:2003rj},
which is consistent with (\ref{abel_inst}).

The derivation of the $U(1)$ instanton partition function easily generalizes to
 non-hyperkahler $\Gamma_{p,q}\subset U(2)$ quotients.
The resulting partition function should correspond to the sum of characters of a conjectural 
chiral algebra defined by the intersection matrix. This however cannot be interpreted in terms of
Kac-Moody algebrae since McKay correspondence does not apply.

\subsection{Central charges}
  Let us compute the central charge of the corresponding two-dimensional CFT
  from the anomaly polynomial of the six-dimensional (2,0) theory
   \cite{Bonelli:2009zp, Alday:2009qq, Nishioka:2011jk}. 
  Consider $N$ M5-branes compactified on a four-manifold $X$. 
  Under a suitable twist,
  the integrated anomaly eight-form over $X$ leads to the central charge of the remaining two-dimensional theory:
    \bea
    c
     =     N \chi(X) + N(N^{2} - 1) (P_{1}(X) + 2 \chi(X)). \label{ciccio}
    \eea
  where $\chi(X) = \int_{X} e(X)$ and $P_{1}(X) = \int_{X} p_{1}(X)$ 
  are the equivariant Euler number and the integrated first Pontryagin class respectively.
  
  Let us focus on $X=X_{p,q}$, namely the blow up of a toric singularity $\mathbb{C}^{2}/\Gamma_{p,q}$.
  Since we are working with the deformation $\epsilon_{1,2}$, 
  $\chi(X_{p,q})$ and $P_{1}(X_{p,q})$ should be computed in the equivariant sense. 
  This gives the following results
    \bea
    \chi(X_{p,q})
     =     \sum_{\ell=0}^{L-1} 1 = L, ~~~~
    P_{1}(X_{p,q})
     =     \sum_{\ell=0}^{L-1} \frac{(\epsilon_{1}^{(\ell)})^{2} 
         + (\epsilon_{2}^{(\ell)})^{2}}{\epsilon_{1}^{(\ell)}\epsilon_{2}^{(\ell)}},
    \eea
  where $L$ is the number of the fixed points of the torus action 
  and $\epsilon_{1,2}^{(\ell)}$ are given in (\ref{w-bis}).
  By plugging these values into \eqref{ciccio} we recover the central charge of the algebra \eqref{copies}.
  The central charge can be easily expressed through $N_{G}$ defined in \eqref{NG} as
    \bea
    c
     =     NL + N(N^{2}-1) \left( \frac{Q^{2}}{p} - N_{G} \right).
    \eea
  Note that since $\epsilon_{1}$ and $\epsilon_{2}$ enter symmetrically in the above formula,
  also the coefficient $N_{G}$ is also symmetric under the exchange $q \leftrightarrow q_{L-1}$,
  following from the argument just below \eqref{xy}. 

\subsubsection*{$A_{p-1}$ ALE space}
  In the case of the blow up of $\mathbb{C}^{2}/\mathbb{Z}_{p}$ ($q = p-1$), 
  the number of the fixed points is $p$ and
  the $\epsilon_{1,2}^{(\ell)}$ are given in (\ref{ale}).
  Thus, we get 
     \bea
    c
     =     N p + \frac{N(N^{2} - 1) Q^{2}}{p},
           \label{centralale}
    \eea
  where we defined $Q = b + 1/b$ with $\sqrt{\epsilon_{1}/\epsilon_{2}} = b$.
  One can check that this is indeed the sum of the central charges of 
  the Heisenberg, affine $\hat{su}(p)_{N}$ and $p$-th para-$W_{N}$ algebras,
  as found in \cite{Nishioka:2011jk}:
    \bea
    c
     =     1 + \frac{N(p^{2} - 1)}{N+p} + \left( \frac{p(N^{2} - 1)}{N+ p} + \frac{N (N^{2} - 1) Q^{2}}{p} \right).
    \eea

\subsubsection*{$\CO_{\mathbb{P}^{1}}(-p)$ space}
  In the case of the $\CO_{\mathbb{P}^{1}}(-p)$ space ($q = 1$), 
  the number of the fixed points is $2$ and
  the $\epsilon_{1,2}^{(\ell)}$ are given in (\ref{o-p}).
  Therefore, we obtain
    \bea
    c
     =     2N + \frac{N(N^{2}-1)}{p} \left( Q^{2} - (p-2)^{2} \right).
    \eea
  In the case of $p=2$, this coincides with (\ref{centralale}) with $p=2$.

\subsubsection*{Other examples}
  For example, we can compute the first cases which are not of the two types above.
  For $\Gamma_{5,2}$ and $\Gamma_{5,3}$, we get, due to the $q \leftrightarrow p_{L-1}$ symmetry, the same result
    \bea
    c
     =     3N + \frac{N (N^{2} - 1)}{5} \left( Q^{2} - 2 \right).
    \eea
  For $\Gamma_{7,2}$, we get
    \bea
    c
     =     3N + \frac{N (N^{2} - 1)}{7} \left( Q^{2} - 8 \right).
    \eea

\section{Conclusions and open issues}
\label{sec:conclusion}

In this paper we started a systematic study of ${\cal N}=2$ gauge theories on toric singularities.
The results we obtained raise a number of issues that it would be interesting to further study.

First of all, the blow-up formula \eqref{full} together with AGT correspondence implies
that the instanton partition function provides an analytic form for the conformal blocks of the
algebra ${\cal A}_N(X_{p,q})$ we discussed in section \ref{sec:CFT}, see \eqref{copies}. Moreover, the one-loop
contribution should be interpreted as the three-point functions of the corresponding two-dimensional CFT.
These should be regarded as the structure constants of the OPE algebra of primaries,
whose spectrum is not known at the moment for general $X_{p,q}$ varieties. It would be interesting
to investigate further this issue. In the ALE case a natural candidate has been proposed
in \cite{Nishioka:2011jk} as the $\zet_p$ parafermionic $W_N$ theory, whose conformal blocks and three-point functions
are however not known in general. Our result can be regarded as a proposal for their computation.
Let us remark that for $p=N=2$ this is verified both for the conformal blocks 
\cite{Belavin:2011pp,Bonelli:2011jx,Ito:2011mw,Belavin:2011sw}
and the three-point functions \cite{Bonelli:2011kv}
and for $p=4$, $N=2$ case in \cite{Wyllard:2011mn, Alfimov:2011ju}.
In the ALE case and $N=2$ one might compare our one-loop formulas with the parafermionic Liouville three-point functions as given in \cite{Bershtein:2010wz}.

Another interesting aspect that could help in elucidating the above problem is to investigate the ${\cal N}=4$ limit
of the partition function in the generic $\Gamma_{p,q}$ case, which should be expressed in terms of the characters of the full conformal algebra
\cite{Bonelli:2011kv}. 
It could be useful in this respect to study the level-rank duality properties of these partition functions
in full generality. 
Let us remark that the algebraic structure that we find in the ALE case contains as a factor the Kac-Moody algebrae 
found in \cite{Nakajima} and further discussed in \cite{Vafa:1994tf, Dijkgraaf:2007sw, Dijkgraaf:2007fe, Tan:2008wp}.


Although we did not discuss two-dimensional CFT correlators, we expect that these would arise by computing
the ${\cal N}=2$ partition function on $S^4/\Gamma_{p,q}$ along the lines of \cite{Pestun:2007rz}. 
It would be also interesting to study BPS observables in the gauge theory on toric singularities, in particular
surface operators and their possible relationship with degenerate fields insertions in the two-dimensional CFT. 
   

The results presented in this paper are obtained under some natural quantum field theoretic assumptions,
namely the patch-by-patch factorization of the full partition function. This is a well established result
for some particular cases, and it would be relevant to have a rigorous proof in general.
This would amount to have a direct computation of the $\ell$-factor in \eqref{ellfact} from the localization
formula in each topological sector of the theory. 
To this end a generalization of the ADHM of \cite{KNakajima} to $\Gamma_{p,q}$ orbifolds would also be welcome.
This could also help to extend the rigorous proof of the AGT correspondence of \cite{SV} to instanton
counting on toric singularities.

An open problem is the geometric engineering of these four-dimensional gauge theories, which would shed light on the
topological string description of their Nekrasov partition function at least in the unrefined $\epsilon_1+\epsilon_2=0$ case.
This issue would naturally lead to a comparison with the results of \cite{Krefl:2011aa}
where the holomorphic anomaly equation for the $A_{1}$ ALE space is discussed.

Finally, the r\^ole of integrable systems in this whole story has to be elucidated. Our result \eqref{full}
again suggests the emergence of a quiver integrable system made of nested copies of the systems relevant for the flat space, whose intertwining
is dictated by the intersection matrix of the $X_{p,q}$ variety.
For example for linear quivers we expect to obtain nested copies of the Calogero-Sutherland system, which is the
relevant one in the flat space case \cite{Estienne:2011qk,SV}.
A more general view on the problem would be gained in terms of a suitable generalization of the Hitchin system
approach to $X_{p,q}$ varieties.

\section*{Acknowledgments}
We would like to thank 
Mikhail Alfimov, 
Francesco Benini,
Hirotaka Irie,
Can Kozcaz, 
Noppadol Mekareeya, 
Takuya Okuda, 
Sara Pasquetti,
Raoul Santachiara, 
Francesco Sala,
Olivier Schiffmann,
Yuji Tachikawa, 
Masato Taki 
and
Meng-Chwan Tan
for useful discussions and comments, 
and the participants to the 5th Workshop on Geometric Methods in Theoretical Physics at SISSA 
for stimulating comments.
In particular we thank Sara Pasquetti for a crucial observation on the title of an earlier version of the draft.
This research was partly supported by the INFN Research Projects PI14, ``Nonperturbative dynamics of gauge theory" 
and TV12, and by   PRIN    ``Geometria delle variet\`a algebriche".

\appendix

\section*{Appendix}

\section{Hirzebruch-Jung resolution: }
\label{HJ}
  The $(p,q)$ toric singularity is the quotient $\complex^2/\Gamma_{p,q}$, 
  where the action is in local coordinates $z_1\to e^{2\pi i/p}z_1$ and $z_2\to e^{2\pi i q/p}z_2$, 
  with $(p,q)$ being coprime and $q<p$. 

The Hirzebruch-Jung resolution of the above singularity is prescribed as follows.
Let $p/q= [e_1, \ldots, e_{L-1}] \equiv e_1-\frac{1}{e_2-\frac{1}{e_3-\ldots}}$
be the continuous fraction expansion of the ratio $p/q$
in terms of the finite sequence of positive integers $\{e_1,\ldots,e_{L-1}\}$.
The fan of the resolved toric variety is given in terms of 
the set of vectors $\left\{v_\ell\right\}_{\ell=0,\ldots,L}$ in $\zet^2$ satisfying the recursion relation
$v_{\ell+1}+v_{\ell-1}=e_\ell v_\ell$ with boundary conditions $v_0=(0,1)$ and $v_L=(p,-q)$.
Each internal vector $v_\ell$, $\ell=1,\ldots, L-1$ of the fan corresponds to a blown-up sphere $\Delta_\ell$. 
Recalling that the intersection pairing of these spheres can be computed from the toric fan
as $\Delta_\ell \cdot \Delta_\ell = v_{\ell-1}\wedge v_{\ell+1}$ 
and $\Delta_\ell \cdot \Delta_{\ell+1}=v_\ell \wedge v_{\ell+1}$, 
we can compute the intersection matrix to be
\be
{\cal I}_{p,q}\equiv -C=
\begin{pmatrix}
-e_1 & 1      &     0   & \dots & 0  \\
1   & - e_2    & \ddots  & {}    & {} \\
0   & \ddots & \ddots  & \ddots& {} \\ 
\vdots  & {}     & \ddots  & \ddots& 1  \\
0   & {}         & {} & 1 & - e_{L-1}
\end{pmatrix}
\label{C}
\ee
The cones of the fan are $\sigma_\ell=\{v_\ell,v_{\ell+1}\}$, with $\ell=0,\ldots, L-1$, and those of the the dual fan
$\sigma_\ell^\star=\{- v_\ell \tau_2 , v_{\ell+1} \tau_2\}$, where 
$\tau_2=${\tiny $\begin{pmatrix} 0 & 1\\ -1 & 0 \end{pmatrix}$}.
The ring of holomorphic functions on the toric variety can be described in terms of the dual cones
as follows.
Let us denote the entries of the vectors of the dual fan as
$\sigma_\ell^\star =\left\{ ( s_\ell,t_\ell ),(\tilde s_\ell ,\tilde t_\ell )\right\}$.
By using this notation, the polynomial ring reads
\be
\oplus_{\ell=0}^{L-1} \complex[w_1^{s_\ell}w_2^{t_\ell},
w_1^{\tilde s_\ell}w_2^{\tilde t_\ell}]
=
\oplus_{\ell=0}^{L-1} \complex[z_1^{ps_\ell-qt_\ell}z_2^{t_\ell},
z_1^{p\tilde s_\ell-q\tilde t_\ell}z_2^{\tilde t_\ell}],
\label{t}
\ee
where we used the invariant variables $w_1=z_1^p$ and $w_2=z_2/z_1^q$.

The torus action $(\complex^\ast)^2$ on the singular variety $(z_1,z_2)\to (e^{i\epsilon_1} z_1, e^{i\epsilon_2} z_2)$
descends to a torus action on the resolved variety whose weights can be  
obtained directly from (\ref{t}) as
\bea
\epsilon_1^{(\ell)} &=& (ps_\ell - q t_\ell)\epsilon_1 + t_\ell \epsilon_2,
\nonumber \\
\epsilon_2^{(\ell)} &=& (p\tilde s_\ell - q \tilde t_\ell)\epsilon_1 + \tilde t_\ell \epsilon_2.
\label{w}
\eea
Let us now consider some explicit examples that are relevant for the main text calculations.

\subsection{$\mathcal{O}_{\mathbb{P}^1}(-p)$ space}

The resolution of the $\Gamma_{p,1}$ singularity is given by the total space of $\mathcal{O}_{\mathbb{P}^1}(-p)$-bundle.
The cones are 
\be
\sigma_0 = \{(0,1),(1,0)\},\quad \ \ \sigma_1 = \{(1,0),(p,-1)\},
\ee
and the dual cones
\be
\sigma_0^\star = \{(1,0),(0,1)\}, \quad \ \ \sigma_1^\star = \{(0,-1),(1,p)\}.
\ee
 The weights of the torus action in the two patches are
\bea
\epsilon^{(0)}_1 &=& p\epsilon_1,\quad  \epsilon^{(0)}_2 =  \epsilon_2 - \epsilon_1, \nonumber\\
\epsilon^{(1)}_1 &=& \epsilon_1-\epsilon_2,\quad  \epsilon^{(1)}_2 = p\epsilon_2. 
\label{o-p}
\eea

\subsection{$A_{p-1}$ ALE space}

The cones of the resolution of the $\Gamma_{p,p-1}$ singularity are
\be
\sigma_\ell = \{(\ell,1-\ell),(\ell+1, -\ell )\} 
\ee
with $\ell=0,\ldots,p-1$. The dual cones are
 \be
\sigma_\ell^\star = \{(1- \ell, -\ell),(\ell, \ell+1 )\} .
\ee
 The weights of the torus action in the $p$ patches are
\be
\epsilon^{(\ell)}_1 = (p -\ell) \epsilon_1 - \ell \epsilon_2, \quad  
\epsilon^{(\ell)}_2 =  (-p + \ell +1)\epsilon_1 + (\ell+1) \epsilon_2.
\label{ale}
\ee

\subsection{$\Gamma_{5,2}$, $\Gamma_{5,3}$ and $\Gamma_{7,2}$ singularities}

Let us consider other simple cases.
For the $\Gamma_{5,2}$ case, the cones are
\be
\sigma_0 = \{(0,1),(1,0)\}, \quad \ \ \sigma_1 = \{(1,0),(3,-1)\},\quad \ \ \sigma_2 = \{(3,-1),(5,-2) \},
\ee
and the dual cones
\be
\sigma_0^\star = \{(1,0),(0,1)\}, \quad \ \ \sigma_1^\star = \{(0,-1),(1,3)\}, \quad \ \ \sigma_2^\star = \{(-1,-3),(2,5) \}.
\ee
 The weights of the torus action in the three patches are
\bea
\epsilon^{(0)}_1 &=& 5\epsilon_1,\quad  \epsilon^{(0)}_2 =  \epsilon_2 - 2 \epsilon_1, \nonumber\\
\epsilon^{(1)}_1 &=& 2 \epsilon_1-\epsilon_2, \quad  \epsilon^{(1)}_2 = -\epsilon_1 + 3\epsilon_2, \nonumber\\
\epsilon^{(2)}_1 &=& \epsilon_1 -3 \epsilon_2,\quad  \epsilon^{(2)}_2 =  5 \epsilon_2. 
\label{5-2}
\eea
For the $\Gamma_{5,3}$ case, the weights of the torus action in the three patches are
\bea
\epsilon^{(0)}_1 &=& 5\epsilon_1,\quad  \epsilon^{(0)}_2 =  \epsilon_2 - 3 \epsilon_1, \nonumber\\
\epsilon^{(1)}_1 &=& 3 \epsilon_1-\epsilon_2, \quad  \epsilon^{(1)}_2 = -\epsilon_1 + 2\epsilon_2, \nonumber\\
\epsilon^{(2)}_1 &=& \epsilon_1 -2 \epsilon_2, \quad  \epsilon^{(2)}_2 =  5 \epsilon_2. 
\label{5-3}
\eea
Finally, for the $\Gamma_{7,2}$ case, we get the weights of the torus action again in the three patches:
\bea
\epsilon^{(0)}_1 &=& 7\epsilon_1,\quad  \epsilon^{(0)}_2 =  \epsilon_2 - 2 \epsilon_1, \nonumber\\
\epsilon^{(1)}_1 &=& 2 \epsilon_1-\epsilon_2, \quad  \epsilon^{(1)}_2 = -\epsilon_1 + 4\epsilon_2, \nonumber\\
\epsilon^{(2)}_1 &=& \epsilon_1 -4 \epsilon_2, \quad  \epsilon^{(2)}_2 =  7 \epsilon_2. 
\label{7-2}
\eea

\subsection{Comment}
In this section, we show a way to solve the relation $v_{\ell+1}+v_{\ell-1}=e_\ell v_\ell$ explicitly.
We define the following two sets of sequences of integers:
\bea
&&p_0 = 0, \qquad
p_1 = 1, \qquad
(p_2=e_1), \qquad
p_{\ell+1} = e_{\ell} p_{\ell} - p_{\ell-1} \quad (\ell \ge 1),
\label{def_pl}
\\
&&q_0 = -1, \qquad
q_1 = 0, \qquad
(q_2=1), \qquad
q_{\ell+1} = e_{\ell} q_{\ell} - q_{\ell-1} \quad (\ell \ge 1).
\eea
By definition, they satisfy the same recursion relation as $v_{\ell}$.
It is known that $p_{\ell}$ and $q_{\ell}$ are coprime to each other
and that they are the numerator and the denominator of
$[e_1, \cdots, e_{\ell-1}]$ respectively:
\bea
[e_1, \cdots, e_{\ell-1}] = \frac{p_{\ell}}{q_{\ell}}.
\qquad (\ell \ge 2)
\eea
In particular, this means that $p_{L}=p$ and $q_{L}=q$.
It is also known that 
\bea
p_{\ell} q_{\ell+1} - q_{\ell} p_{\ell+1} = 1.
\label{pqqp} 
\eea

Comparing the recursion relation and the boundary condition,
we identify
\bea
s_{\ell} = - q_{\ell},
\quad
\tilde{s}_{\ell} = q_{\ell+1},
\quad
t_{\ell} = - p_{\ell},
\quad
\tilde{t}_{\ell} = p_{\ell+1}.
\label{19}
\eea
In terms of \eqref{19} we have 
\bea
\epsilon_1^{(\ell)} &=& -(pq_\ell - q p_\ell)\epsilon_1 - p_\ell \epsilon_2, \nonumber \\
\epsilon_2^{(\ell)} &=& (pq_{\ell+1} - q p_{\ell+1})\epsilon_1 + p_{\ell+1} \epsilon_2.
\label{w-bis}
\eea
Explicitly,
\bea
\epsilon_1^{(0)} = p\epsilon_1, &&\quad 
\epsilon_2^{(0)} = - q\epsilon_1 + \epsilon_2, 
\nonumber \\
\epsilon_1^{(1)} = q\epsilon_1 - \epsilon_2, &&\quad 
\epsilon_2^{(1)} = (p - q e_1 ) \epsilon_1 + e_1 \epsilon_2, 
\nonumber \\
\epsilon_1^{(2)} = (- p + q e_1 ) \epsilon_1 - e_1 \epsilon_2, && \quad 
\epsilon_2^{(2)} = \cdots, 
\nonumber \\
&&\cdots
\nonumber \\
\epsilon_1^{(L-2)} = \cdots && \quad 
\epsilon_2^{(L-2)} = - \epsilon_1 + p_{L-1} \epsilon_2, 
\nonumber \\
\epsilon_1^{(L-1)} = \epsilon_1 - p_{L-1} \epsilon_2 && \quad 
\epsilon_2^{(L-1)} = p \epsilon_2,
\label{several_omega}
\eea
where we have used (\ref{pqqp}) for $\ell=L-1$ to calculate 
$\epsilon_2^{(L-2)}$ and $\epsilon_1^{(L-1)}$.

\section{Various summations of Omega deformation parameters}
\label{sec:sum}
By construction, the Omega deformation parameters $\epsilon_1^{(\ell)}$ and $\epsilon_2^{(\ell)}$
are related by $\epsilon_1^{(\ell)} = - \epsilon_2^{(\ell-1)}$, 
and respectively satisfy the recursion relations
\bea
\epsilon_1^{(\ell-1)} + \epsilon_1^{(\ell+1)} = e_{\ell} \epsilon_1^{(\ell)},
\qquad
\epsilon_2^{(\ell-1)} + \epsilon_2^{(\ell+1)} = e_{\ell+1} \epsilon_2^{(\ell)}.
\label{rec_eps}
\eea
They satisfy the boundary condition
 $\epsilon_1^{(0)}=p\epsilon_1$ and 
$\epsilon_2^{(L-1)}=p\epsilon_2$.

It follows from (\ref{pqqp}) that 
\bea
p_{\ell+1} \epsilon_1^{(\ell)} 
+ p_{\ell} \epsilon_2^{(\ell)} = \epsilon_1^{(0)}.
\eea
By using this identity and \eqref{w-bis}, we can show that
\bea
\sum_{\ell=0}^{m} \frac{1}{\epsilon_1^{(\ell)}\epsilon_2^{(\ell)} } 
= \frac{p_m}{\epsilon_1^{(0)} \epsilon_2^{(m)}},
\eea
which for $m=L-1$ is
\bea
\sum_{\ell=0}^{L-1} \frac{1}{\epsilon_1^{(\ell)}\epsilon_2^{(\ell)} } 
= \frac{p}{\epsilon_1^{(0)} \epsilon_2^{(L-1)}}
= \frac{1}{p \epsilon_1 \epsilon_2}.
\label{eps-F0}
\eea

We list here other useful formulae used in section \ref{sec:gauge}:
  \bea
   \sum_{\ell=0}^{L-1} 
   \frac{\epsilon_1^{(\ell)} + \epsilon_2^{(\ell)}}{\epsilon_1^{(\ell)}\epsilon_2^{(\ell)}}
     &=&  \frac{\epsilon_1+\epsilon_2}{p \epsilon_1 \epsilon_2},
   \label{eps-H}
   \\
   \sum_{\ell=0}^{L-1}
   \frac{\epsilon_1^{(\ell)} k^{(\ell+1)}_{\alpha} + \epsilon_2^{(\ell)} k^{(\ell)}_{\alpha}}%
       {\epsilon_1^{(\ell)}\epsilon_2^{(\ell)}}
     &=& 0,
   \label{eps-pF0}
   \\
   \sum_{\ell=0}^{L-1}
   \frac{(\epsilon_1^{(\ell)} + \epsilon_2^{(\ell)})^2}{\epsilon_1^{(\ell)}\epsilon_2^{(\ell)}}
     &=& \frac{(\epsilon_1+\epsilon_2)^2}{p\epsilon_1 \epsilon_2}
         - \frac{p_{L-1}+q+2}{p} + 2L - \sum_{\ell=1}^{L-1} e_{\ell},
   \label{eps-G}
   \\
     \Bigl( 
     &=& \frac{(\epsilon_1+\epsilon_2)^2}{p\epsilon_1 \epsilon_2}
         + \frac{2p - 2 - p_{L-1} - q}{p} 
         + \sum_{\ell=1}^{L-1} (2 - e_{\ell})
     \Bigr)
   \nonumber \\
   \sum_{\ell=0}^{L-1}
   \frac{(\epsilon_1^{(\ell)} + \epsilon_2^{(\ell)})
        (\epsilon_1^{(\ell)} k^{(\ell+1)}_{\alpha} + \epsilon_2^{(\ell)} k^{(\ell)}_{\alpha} )}%
        {\epsilon_1^{(\ell)} \epsilon_2^{(\ell)} } 
     &=& \sum_{\ell=1}^{L-1} (2 - e_{\ell}) k_{\alpha}^{(\ell)},
     \label{eps-pH} \\
     \Bigl( 
     &=& k^{(1)}_{\alpha} + k^{(L-1)}_{\alpha} 
             - \sum_{\ell=1}^{L-1} \sum_{m=1}^{L-1} 
             C_{\ell m} k^{(m)}_{\alpha} 
     \Bigr)
   \nonumber \\
   \sum_{\ell=0}^{L-1} 
   \frac{ (\epsilon_1^{(\ell)} k^{(\ell+1)}_{\alpha} 
       + \epsilon_2^{(\ell)} k^{(\ell)}_{\alpha} ) 
       (\epsilon_1^{(\ell)} k^{(\ell+1)}_{\beta} 
       + \epsilon_2^{(\ell)} k^{(\ell)}_{\beta} ) }%
       {\epsilon_1^{(\ell)}\epsilon_2^{(\ell)}}
     &=& - \sum_{\ell=1}^{p-1} \sum_{m=1}^{p-1} 
         k^{(\ell)}_{\alpha} C_{\ell m} k^{(m)}_{\beta},
   \label{eps-tau}
   \eea
 Note that (\ref{eps-G}) respect the symmetry exchanging
 \bea
 p_{L-1} = \frac{p}{[e_{L-1}, \cdots, e_1]}
 \quad {\rm and} \quad 
 q = \frac{p}{[e_{1}, \cdots, e_{L-1}]}.
 \eea

\paragraph{Derivation of \eqref{shift} from \eqref{xy}}

 From \eqref{xy} we have 
 \bea
 a^{(\ell)}_{\alpha} 
     &=& a_{\alpha} 
       - \sum_{\ell=1}^{L-1} (p q_{\ell} - q p_{\ell}) u_{(\ell)\alpha}\epsilon_1 
       - \sum_{m=1}^{\ell} u_{(m)\alpha} \epsilon_1^{(m)}
   =a_{\alpha} 
       - \sum_{m=0}^{\ell} u_{(m)\alpha} \epsilon_1^{(m)},
 \label{weight_u}
 \eea 
where we have defined
 \bea
 u_{(0)\alpha} \equiv \sum_{\ell=1}^{L-1} \frac{p q_{\ell} - q p_{\ell}}{p} u_{(\ell)\alpha}.
 \label{def_u0}
 \eea
Now, we are going to rewrite it in terms of $k=C^{-1}u$. 
Note that
 \bea
  pq_{\ell}-qp_{\ell} 
    = - p \left( C^{-1} \right )^{1 \ell}
 \eea
for $ 1 \le \ell \le L-1$.
Therefore, (\ref{def_u0}) is rewritten as
 \bea
 u_{(0)\alpha}
      = \sum_{\ell=1}^{L-1} \frac{p q_{\ell} - q p_{\ell}}{p} u_{(\ell)\alpha}
      = - \sum_{\ell=1}^{L-1} \left( C^{-1} \right )^{1 \ell} u_{(\ell)\alpha}
      = - k^{(1)}_{\alpha}
 \eea
and thus, (\ref{weight_u}) is rewritten as
 \bea
 a^{(\ell)}_{\alpha} 
     = a_{\alpha} 
       + k^{(1)}_{\alpha} \epsilon_1^{(0)} 
       - \sum_{m=1}^{\ell} u_{(m)\alpha} \epsilon_1^{(m)}.
         \label{aellalpha}
 \eea  
By considering the third term and taking into account that $k^{(0)}=k^{(L)}=0$,
we find 
\be
- \sum_{m=1}^{\ell} u_{(m)\alpha} \epsilon_1^{(m)}
=
 - k^{(1)}_{\alpha} \epsilon_1^{(0)}      
         + k^{(\ell+1)}_{\alpha} \epsilon_1^{(\ell)} 
         + k^{(\ell)}_{\alpha} \epsilon_2^{(\ell)},
\ee
where we used (\ref{rec_eps}).
Therefore, we finally obtain the shift formula (\ref{shift}).

\section{$U(2)$ SYM theory on $A_2$ ALE}
\label{sec:explicit}

We give the explicit calculation for the $A_2$ ALE instanton partition function.
We expand the instanton partition function \eqref{aleinst} as
\bea
Z_{\rm inst} (\epsilon_1,\epsilon_2,a; \vec{I}; c_1^{(1)}, c_1^{(2)})
= \sum_{c^{(1)}, ~c^{(2)}} Z^{(c^{(1)}, ~c^{(2)})} (\vec{I}) 
\xi_{1}^{c_1^{(1)}} \xi_{2}^{c_1^{(2)}},
\eea
where $a \equiv (a_1-a_2)/2$ and $I_{\alpha} = 3k_{\alpha}^{(1)}$ mod $3$.
We note that $\vec{k}^{(1)}$ and $c_1^{(\ell)}$
are not independent but are related as
\bea
c_1^{(1)} - c_1^{(2)} = - (I_{1}+I_{2})
\qquad
{\rm mod} \,\, 3.
\eea
We calculate the lowest order in instanton number 
for $ -1 \le c_1^{(1)}, c_1^{(2)} \le 1$.
The results are given by
\begin{align}
Z^{(0,0)} (0,0) = 
1 + \cdots
\nonumber 
\end{align}
\begin{align}
&Z^{(0,0)} (1,2) 
\cr
&= 
\left[
\frac{-1}{(2a-\epsilon_2)(2a+\epsilon_1)} 
+ \frac{-1}{(2a+\epsilon_1)(2a+2\epsilon_1+\epsilon_2)}
+ \frac{-1}{(2a-\epsilon_1-2\epsilon_2)(2a-\epsilon_2)}
\right] q^{\frac{2}{3}}
 + \cdots
 \nonumber 
\end{align}
\begin{align}
&Z^{(0,0)} (2,1)
\cr
&= 
\left[
\frac{-1}{(2a+\epsilon_2)(2a-\epsilon_1)} 
+ \frac{-1}{(2a-\epsilon_1)(2a-2\epsilon_1-\epsilon_2)}
+ \frac{-1}{(2a+\epsilon_1+2\epsilon_2)(2a+\epsilon_2)}
\right] q^{\frac{2}{3}}
 + \cdots
 \nonumber 
\end{align}
\begin{align}
Z^{(0,1)} (2,2) = 
\left[
\frac{-1}{(2a-\epsilon_1-\epsilon_2)(2a)} 
+ \frac{-1}{(2a)(2a+\epsilon_1+\epsilon_2)}
\right] q^{\frac{2}{3}}
 + \cdots
 \nonumber 
\end{align}
\begin{align}
Z^{(0,1)} (0,1)= 
q^{\frac{1}{3}} + \cdots
\nonumber 
\end{align}
\begin{align}
Z^{(0,1)} (1,0) = 
q^{\frac{1}{3}} + \cdots
\nonumber 
\end{align}
\begin{align}
Z^{(1,0)} (1,1) = 
\left[
\frac{-1}{(2a-\epsilon_1-\epsilon_2)(2a)} 
+ \frac{-1}{(2a)(2a+\epsilon_1+\epsilon_2)}
\right] q^{\frac{2}{3}}
 + \cdots
 \nonumber 
\end{align}
\begin{align}
Z^{(1,0)} (0,2)= 
q^{\frac{1}{3}} + \cdots
\nonumber 
\end{align}
\begin{align}
Z^{(1,0)} (2,0) = 
q^{\frac{1}{3}} + \cdots
\nonumber 
\end{align}
\begin{align}
Z^{(0,-1)} (1,1) = 
\left[
\frac{-1}{(2a-\epsilon_1-\epsilon_2)(2a)} 
+ \frac{-1}{(2a)(2a+\epsilon_1+\epsilon_2)}
\right] q^{\frac{2}{3}}
+ \cdots
\nonumber 
\end{align}
\begin{align}
Z^{(0,-1)} (0,2) = 
q^{\frac{1}{3}} + \cdots
\nonumber 
\end{align}
\begin{align}
Z^{(0,-1)} (2,0) = 
q^{\frac{1}{3}} + \cdots
\nonumber 
\end{align}
\begin{align}
Z^{(-1,0)} (2,2) = 
\left[
\frac{-1}{(2a-\epsilon_1-\epsilon_2)(2a)} 
+ \frac{-1}{(2a)(2a+\epsilon_1+\epsilon_2)}
\right] q^{\frac{2}{3}}
+ \cdots
\nonumber 
\end{align}
\begin{align}
Z^{(-1,0)} (0,1) = 
q^{\frac{1}{3}} + \cdots
\nonumber 
\end{align}
\begin{align}
Z^{(-1,0)} (1,0) = 
q^{\frac{1}{3}} + \cdots
\nonumber 
\end{align}
\begin{align}
Z^{(1,1)} (0,0) = 
\left[
\frac{-1}{(2a-\epsilon_1-\epsilon_2)(2a)} 
+ \frac{-1}{(2a)(2a+\epsilon_1+\epsilon_2)}
\right] q^{\frac{2}{3}}
\nonumber 
\end{align}
\begin{align}
Z^{(1,1)} (1,2) = 
q + \cdots
\nonumber 
\end{align}
\begin{align}
Z^{(1,1)} (2,1) = 
q + \cdots
\nonumber 
\end{align}
\begin{align}
Z^{(-1,-1)} (0,0) = 
\left[
\frac{-1}{(2a-\epsilon_1-\epsilon_2)(2a)} 
+ \frac{-1}{(2a)(2a+\epsilon_1+\epsilon_2)}
\right] q 
+ \cdots
\nonumber 
\end{align}
\begin{align}
Z^{(-1,-1)} (1,2) = 
q^{\frac{2}{3}}
+ \cdots
\nonumber 
\end{align}
\begin{align}
Z^{(-1,-1)} (2,1) = 
q^{\frac{2}{3}}
+ \cdots
\nonumber 
\end{align}
\begin{align}
Z^{(1,-1)} (2,2) = 
\left[
\frac{-1}{(2a-\epsilon_1-\epsilon_2)(2a)} 
+ \frac{-1}{(2a)(2a+\epsilon_1+\epsilon_2)}
\right]
q^{\frac{2}{3}}
+ \cdots
\nonumber 
\end{align}
\begin{align}
Z^{(1,-1)} (0,1)= q^{\frac{1}{3}} + \cdots
\nonumber 
\end{align}
\begin{align}
Z^{(1,-1)} (1,0)= q^{\frac{1}{3}} + \cdots
\nonumber 
\end{align}

\begin{align}
Z^{(-1,1)} (1,1) = 
\left[
\frac{-1}{(2a-\epsilon_1-\epsilon_2)(2a)} 
+ \frac{-1}{(2a)(2a+\epsilon_1+\epsilon_2)}
\right]
q^{\frac{2}{3}}
+ \cdots
\nonumber 
\end{align}
\begin{align}
Z^{(1,-1)} (0,2)= q^{\frac{1}{3}} + \cdots
\nonumber 
\end{align}
\begin{align}
Z^{(1,-1)} (2,0)= q^{\frac{1}{3}} + \cdots
\nonumber 
\end{align}
These results coincide with the computation done using the orbifold projection method of \cite{Fucito:2004ry}.
 

\bibliographystyle{ytphys}
\small\baselineskip=.97\baselineskip
\bibliography{ref}

\end{document}